\documentclass[%
aip,
jcp,%
amsmath,amssymb,
fullbibliographhy,
reprint,%
]{revtex4-1}
\usepackage[latin1]{inputenc}
\usepackage{amsmath,amsbsy}

\usepackage{amsfonts}
\usepackage{amssymb,bm}
\usepackage{graphicx}
\usepackage{physics}
\usepackage{upgreek}
\usepackage{txfonts}

\DeclareMathAlphabet{\pazocal}{OMS}{zplm}{m}{n}
\newcommand{\pP}{\ensuremath{\pazocal{P}}}
\newcommand{\pQ}{\ensuremath{\pazocal{Q}}}
\newcommand{\pL}{\ensuremath{\pazocal{L}}}
\newcommand{\pK}{\ensuremath{\pazocal{K}}}

\newcommand{\pR}{\ensuremath{\pazocal{R}}}

\newcommand{\el}{\ensuremath{\mathrm{e}}}

\newcommand{\sing}{\ensuremath{\mathrm{S}}}
\newcommand{\trip}{\ensuremath{\mathrm{T}}}
\newcommand{\ipi}{\ensuremath{\mathrm{I}}}
\renewcommand{\op}[1]{\ensuremath{\hat{#1}}}

\newcommand{\sX}{\ensuremath{\mathsf{X}}}
\usepackage{xcolor}
\newcommand{\highlight}[1]{{\color{black}#1}}

\DeclareMathAlphabet{\mathbfcal}{OMS}{cmsy}{b}{n}

\begin{document}
	
\title{Electron spin relaxation in radical pairs: Beyond the Redfield approximation}
\author{Thomas P. Fay}
\affiliation{Department of Chemistry, University of Oxford, Physical and Theoretical Chemistry Laboratory, South Parks Road, Oxford, OX1 3QZ, UK}
\author{Lachlan P. Lindoy}
\affiliation{Department of Chemistry, University of Oxford, Physical and Theoretical Chemistry Laboratory, South Parks Road, Oxford, OX1 3QZ, UK}
\author{David E. Manolopoulos}
\affiliation{Department of Chemistry, University of Oxford, Physical and Theoretical Chemistry Laboratory, South Parks Road, Oxford, OX1 3QZ, UK}

\begin{abstract}
	Relaxation processes can have a large effect on the spin selective electron transfer reactions of radical pairs. These processes are often treated using phenomenological relaxation superoperators, or with some model for the microscopic relaxation mechanism treated within \textcolor{black}{Bloch-Redfield-Wangsness} theory. Here we demonstrate that an alternative perturbative relaxation theory, based on the Nakajima-Zwanzig equation, has certain advantages over Redfield theory. In particular, the Nakajima-Zwanzig equation does not suffer from the severe positivity problem of Redfield theory in the static disorder limit. Combining the Nakajima-Zwanzig approach consistently with the Schulten-Wolynes semiclassical method, we obtain an efficient method for modelling the spin dynamics of radical pairs containing many hyperfine-coupled nuclear spins. This is then used to investigate the spin-dependent electron transfer reactions and intersystem crossing of dimethyljulolidine-naphthalenediimide (DMJ-NDI) radical ion pairs. By comparing our simulations with experimental data, we find evidence for a field-independent contribution to the triplet quantum yields of these reactions which cannot be explained by electron spin relaxation alone.
\end{abstract}

\maketitle
	
\section{Introduction}

Magnetic field effects on the reactions of radical pairs arise due to magnetic interactions between the electron spins and both the applied magnetic field and hyperfine-coupled nuclear spins.\cite{Rodgers2009,Steiner1989} The interplay of the coherent electron spin dynamics and spin state selective radical pair recombination can have dramatic effects on these reactions, \textcolor{black}{and as such magnetic field effects provide a powerful probe of the recombinative electron transfer and intersystem crossing processes in a wide range of systems. Examples include systems being engineered to mimic biochemical reactions,\cite{Wasielewski2006, Scott2009a, Scott2011, Zollitsch2018, Hasharoni1995,Kerpal2019} and those being explored as potential molecular spintronic devices and qubits.\cite{Sun2014, Rugg2017, Wu2018}} However, coupling between the electron spins and molecular vibrations and rotations leads to electron spin relaxation, which can also play an important role in the spin dynamics. It is is therefore essential to be able to model this relaxation in order to obtain the correct interpretation of experimental measurements.\cite{Steiner1989,Atherton1993,Goldman2001}

Three main methods have been employed to model electron spin relaxation effects in the past: the Stochastic Liouville equation,\cite{Kubo1963,Freed1971,Vega1975,Lau2010, Vega1975, Pedersen1973b, Pedersen1994} Redfield theory,\cite{Redfield1965,Goldman2001, Nicholas2010} and phenomenological relaxation models.\cite{Kattnig2016,Lukzen2017,Steiner2018} The Stochastic Liouville equation provides an exact treatment of the spin relaxation even when the modulation of the spin interactions is significant and occurs on a slow time-scale.\cite{Lau2010} However calculations with this equation are often prohibitively expensive for all but the smallest spin systems. Bloch-Redfield-Wangsness theory (hereafter referred to simply as Redfield theory) provides an approximate treatment of the relaxation which is valid if the modulation of the spin interactions is weak and occurs on a short time scale, which is often the case for small radicals tumbling in solution.\cite{Redfield1965,Wangsness1953,Goldman2001,Nicholas2010} This method can treat spin relaxation in systems with many more coupled spins than the Stochastic Liouville equation, but it is still quite expensive for radical pairs containing more than \textcolor{black}{10} or so nuclear spins. Furthermore, the commonly used implementation of Redfield theory\cite{Lukzen2017,Kattnig2016a,Steiner2018,Worster2016} is {inconsistent} with the Stochastic Liouville equation in systems in which the singlet and triplet radical pair recombination reactions occur at different rates.\cite{Steiner1989, Haberkorn1976, Ivanov2010, Fay2018} Phenomenological relaxation models use simple additional terms in the radical pair quantum master equation to approximate different types of relaxation, such as spin-spin and spin-lattice relaxation and singlet-triplet dephasing,\cite{Kattnig2016} and are often derived by making additional approximations to Redfield theory. These simple models are often more computationally efficient \textcolor{black}{to employ in simulations}, but in using them a detailed picture of the microscopic origin of the relaxation is lost. \highlight{Many methods that go beyond Redfield theory have been proposed in the open quantum systems literature.\cite{Breuer2001,Berkelbach2012,Chen2016,Neufeld2003} However, insofar as we are aware, none of these has been applied in the context of radical pair spin dynamics as we shall do here.}

In modelling magnetic field effects in real radical pair reactions, it is desirable to find a method that is accurate, consistent with the Stochastic Liouville equation, simple and efficient to implement for realistically large spin systems, and compatible with a detailed microscopic model of the molecular motion of the radical pair. (Computational efficiency is particularly important because radical pair models often contain several unknown parameters which must be fit to experimental data, a process that typically involves performing many consecutive simulations.) In this paper we propose such a method, and illustrate its utility with an example application to the spin dynamics of dimethyljulolidine-naphthalenediimide (DMJ-NDI) radical pairs with a total of 25 hyperfine-coupled nuclear spins. We also show that the method is accurate over the full range of time scales of molecular motion, from the extreme narrowing limit (fast motion) to the static disorder limit (slow motion).

We begin in Sec.~II by comparing the Redfield\cite{Redfield1965,Goldman2001,Nicholas2010,Wangsness1953} and Nakajima-Zwanzig\cite{Nakajima1958,Zwanzig1960,Wang1976,Nishijima1957} approaches to treating electron spin relaxation. We show that Nakajima-Zwanzig theory, which has largely been overlooked in previous studies of spin relaxation in radical pairs, has certain advantages over the more commonly used Redfield theory. We then propose a method that consistently combines Nakajima-Zwanzig theory with the widely used Schulten-Wolynes semiclassical approximation to the spin dynamics.\cite{Schulten1978, Lawrence2016, Hoang2018, Miura2010} The efficiency of the resulting combination allows us to explore in detail the effect of various relaxation mechanisms on the recombination reactions of DMJ-NDI radical pairs, magnetic field effects on which have  been studied experimentally by Scott \textit{et al.}\cite{Scott2009a} This we do in Secs.~III to V. An interesting observation in some radical pair systems,\cite{Fay2017,Lukzen2017,Klein2015} including those explored here, is that an additional triplet product formation pathway is needed to explain the observed magnetic field effects on triplet product yields. Using the method proposed here, we investigate several possible mechanisms for this additional triplet product formation, as well as the role of various plausible electron spin relaxation processes. Our conclusions are presented in Sec.~VI.

\section{Theory}

The coherent intersystem crossing and spin-selective electron transfer reactions of radical pairs are described using the spin density operator of the radical pair state, $\op{\rho}(t)$. Spin interactions in the radical pairs fluctuate due to modulation by molecular motion, and as such the density operator must be averaged over all realisations of these fluctuations, which we denote with $\ev{\cdots}$.\cite{Goldman2001,Atherton1993} From the ensemble average spin density operator $\ev{\op{\rho}(t)}$, all observables of the radical pair spin system can be calculated, such as the singlet and triplet quantum yields 
\begin{subequations}\label{yields-eq}
	\begin{align}
	\Phi_\sing &= k_\sing \int_0^\infty \Tr[\op{P}_\sing \ev{\op{\rho}(t)}] \dd{t}\\
	\Phi_\trip &= k_\trip \int_0^\infty \Tr[\op{P}_\trip \ev{\op{\rho}(t)}] \dd{t},
	\end{align}
\end{subequations}
and the radical pair lifetime
\begin{align}\label{rplifetime-eq}
\tau_\mathrm{RP} = \int_0^\infty \Tr[\ev{\op{\rho}(t)}] \dd{t}.
\end{align}
Here $\op{P}_\sing$ and $\op{P}_\trip$ are projection operators onto the singlet and triplet electron spin states, and $k_\sing$ and $k_\trip$ are the singlet and triplet recombination rate constants.

When molecular motion modulates the spin interactions, the spin density operator is a function of both time, $t$, and the nuclear configuration of the radical pair, $\mathsf{X}$, and $\ev{\op{\rho}(t)} = \int\dd{\sX} \op{\rho}(t,\sX)$. Here $\mathsf{X}$ denotes a generalised set of coordinates describing the nuclear motion, the precise definition of which depends on the chosen model for the motion of the radical pair. For example, in the case of a rigid molecular radical pair $\mathsf{X}$ is the orientation of the molecule specified by the set of three Euler angles, $\mathsf{X}=\Omega=(\phi,\theta,\psi)$. The evolution of $\op{\rho}(t,\mathsf{X})$ can be modelled using \textcolor{black}{the Stochastic Liouville equation\cite{Kubo1963,Freed1971,Vega1975,Lau2010, Pedersen1973b, Pedersen1994} with the Haberkorn reaction term\cite{Haberkorn1976,Ivanov2010,Fay2018}}
\begin{align}
\pdv{t}\op{\rho}(t,\mathsf{X}) = -\frac{i}{\hbar}\left[\op{H}(\mathsf{X}),\op{\rho}(t)\right]-\left\{\op{K},\op{\rho}(t,\mathsf{X})\right\}+\mathsf{D}\,\op{\rho}(t,\mathsf{X}),\label{haber-eq}
\end{align}
where $[\cdot,\cdot]$ and $\{\cdot,\cdot\}$ denote a commutator and an anti-commutator. Here $\mathsf{D}$ is the operator describing the Stochastic evolution of the nuclear coordinate distribution (e.g., for a rigid molecular radical pair where $\mathsf{X}=\Omega$, $\mathsf{D}$ is the rotational diffusion operator). $\op{K}$ is the Haberkorn reaction operator,\cite{Haberkorn1976,Ivanov2010,Fay2018} which accounts for the spin-selective electron transfer reactions to singlet and triplet product states
\begin{align}
\op{K} = \frac{k_\sing}{2}\op{P}_\sing + \frac{k_\trip}{2}\op{P}_\trip,
\end{align}
and $\op{H}(\mathsf{X})$ is the spin Hamiltonian for the radical pair,  which can be written as a sum of two single radical Hamiltonians and an electron spin coupling term
\begin{align}
\op{H}(\mathsf{X}) = \op{H}_1(\mathsf{X}) +\op{H}_2(\mathsf{X}) + \op{H}_{12}(\mathsf{X}).
\end{align}
The single radical parts, $\op{H}_i(\sX)$, each involve an electron Zeeman interaction term and electron-nuclear spin coupling terms for the $N_i$ nuclear spins in the radical,
\begin{align}
\op{H}_i(\sX) = \mu_\mathrm{B}\op{\vb{S}}_i\cdot\vb{g}_i(\sX)\cdot\vb{B}_0 + \sum_{k=1}^{N_i}\op{\vb{S}}_i\cdot\vb{A}_{ik}(\sX)\cdot\op{\vb{I}}_{ik}.
\end{align}
Here $\op{\vb{S}}_i$ and $\op{\vb{I}}_{ik}$ are the lab frame unitless electron and nuclear spin vector operators for radical $i$, $\vb{g}_i(\sX)$ is the g-tensor for the radical, $\vb{A}_{ik}(\sX)$ is the hyperfine coupling tensor between the electron spin and nuclear spin $k$, $\vb{B}_0$ is the external magnetic field and $\mu_\mathrm{B}$ is the Bohr magneton. The electron spin coupling part can be written as a sum of a scalar coupling term and a dipolar coupling term,
\begin{align}
\op{H}_{12}(\sX) = -2 J(\sX)\op{\vb{S}}_1\cdot\op{\vb{S}}_2 + \op{\vb{S}}_1\cdot \vb{D}(\sX)\cdot\op{\vb{S}}_2,
\end{align}
where $J(\sX)$ is the scalar coupling and $\vb{D}(\sX)$ is the dipolar coupling tensor. 

The spin coupling terms, $\vb{g}_i(\sX)$, $\vb{A}_{ik}(\sX)$, $J(\sX)$ and $\vb{D}(\sX)$ all in general depend on the nuclear configuration of the radical pair, for example the orientation of the radicals relative to the lab frame.\cite{Nicholas2010,Kattnig2016, Atherton1993} The fluctuations in the spin coupling parameters give rise to various types of spin relaxation. In the following sections we will describe how to treat the effect of these fluctuations on the dynamics of $\ev{\op{\rho}(t)}$ using the Nakajima-Zwanzig equation. We shall take the initial density operator to be a product of an initial spin density operator, $\ev{\op{\rho}(0)}$, and the equilibrium density, $p_0(\sX)$, of $\sX$,
\begin{align}
\op{\rho}(0,\sX) = \ev{\op{\rho}(0)}p_0(\sX),
\end{align}
in which $\mathsf{D}p_0(\sX) = 0$.

\subsection{Spin relaxation}

Spin relaxation in radical pairs arises from fluctuations in spin-spin interactions due to the thermal motion of the nuclei. The full spin Hamiltonian can in general be split into \textcolor{black}{an ensemble averaged part}, $\op{H}_0$, and a configuration dependent fluctuation part, $\op{V}(\sX)$,
\begin{align}
\op{H}_0 &= \ev{\op{H}} = \int\dd{\sX}p_0(\sX)\op{H}(\sX)  \\
\op{V}(\sX) &= \op{H}(\sX) - \ev{\op{H}}.
\end{align}
The fluctuating part of the Hamiltonian, $\op{V}(\sX)$, can in general be decomposed as 
\begin{align}
\op{V}(\sX) = \sum_{j}f_j(\sX)\,\op{A}_j,
\end{align}
where $f_j(\sX)$ \textcolor{black}{are scalar-valued functions of the nuclear configuration} and $\op{A}_j$ are operators on the spin system. From the definition of the fluctuation term, it follows that the ensemble average of this is zero, $\ev*{\op{V}}=0$. Furthermore, we denote the correlation functions for $f_j(\sX)$ as
\begin{align}
g_{jk}(\tau) = \ev{f_j(0)^*f_k(\tau)} =\int\dd{\sX}f_j(\sX)^*e^{\mathsf{D}\tau}f_k(\sX)p_0(\sX).
\end{align}
Nakajima-Zwanzig theory can be used to obtain a perturbative master equation for the ensemble averaged spin density operator which depends only these correlation functions.

\subsection{Nakajima-Zwanzig theory}

The Nakajima-Zwanzig equation\cite{Nakajima1958,Zwanzig1960} is a formally exact master equation for the projected density operator $\pP\op{\rho}(t,\sX)$,
\begin{align}
\pdv{t}\pP\op{\rho}(t,\sX) = \pP \pL \pP \op{\rho}(t,\sX) + \int_0^t\dd{\tau} \pK(t-\tau) \pP\op{\rho}(\tau,\sX),
\end{align}
where $\pL = -(i/\hbar)[\op{H}(\sX),\cdot]-\{\op{K},\cdot\}+\mathsf{D}$ is the full Liouvillian and we have assumed that $\pP\op{\rho}(0,\sX) = \op{\rho}(0,\sX)$. The kernel $\pK(t)$ is given by
\begin{align}
\pK(t) = \pP \pL \pQ e^{\pQ \pL t} \pQ \pL \pP,
\end{align}
in which $\pQ = 1 - \pP$. For this problem we will define $\pP$ as follows, where $\op{O}(\sX)$ is an arbitrary operator which is also a function of $\sX$,
\begin{align}
\pP \op{O}(\sX) = p_0(\sX) \int\dd{\sX} \op{O}(\sX).
\end{align}
We also define the perturbation Liouvillian as $\pL_V = -(i/\hbar)[\op{V}(\sX),\cdot]$ and the reference Liouvillian as $\pL_0 = \pL - \pL_V$. Expanding the kernel to second order in $\pL_V$ and making the Markovian approximation,\cite{Zwanzig1960,Fay2018,Fay2019} we obtain the following perturbative master equation for $\pP\op{\rho}(t,\sX)$,
\begin{align}
\pdv{t} \pP\op{\rho}(t,\sX) = \left(\pL_0+\pK^{(2)}\right) \pP\op{\rho}(t,\sX),
\end{align}
where
\begin{align}
\pK^{(2)} = \int_0^{\infty}\!\!\!\!\! \dd\tau\, \pP\pL_Ve^{\pL_0\tau}\pL_V\pP.
\end{align}
Here we have used the fact that $\pL_0\pP = \pP\pL_0$ and $\pP\pL_V\pP = 0$. By integrating out $\sX$ we obtain the Nakajima-Zwanzig perturbative master equation for the ensemble averaged density operator,
\begin{align}
\begin{split}
\dv{t}\ev{\op{\rho}(t)} &= -\frac{i}{\hbar}[\op{H}_0,\ev{\op{\rho}(t)}]+\{\op{K},\ev{\op{\rho}(t)}\} +\pR_\mathrm{NZ}\ev{\op{\rho}(t)},
\end{split}
\end{align}
in which the Nakajima-Zwanzig relaxation superoperator, $\pR_\mathrm{NZ}$, is given by
\begin{align}
\pR_\mathrm{NZ} &= \int_0^\infty\!\!\!\!\! \dd{\tau}\int \dd{\sX}{\pL_Ve^{\pL_0\tau}\pL_V}p_0(\sX) \\
&= -\sum_{j,k} \int_0^\infty \!\!\!\!\! \dd{\tau}g_{jk}(\tau)\pazocal{A}_j^\dag e^{\ev{\pL}\tau}\pazocal{A}_k,\label{nz-superop-eq}
\end{align}
with $\pazocal{A}_j^\dag = (1/\hbar)[\op{A}_j^\dag,\cdot]$, $\pazocal{A}_k= (1/\hbar)[\op{A}_k,\cdot]$, and $\ev{\pL} = \pL_0 - \mathsf{D}$. 

\subsection{Redfield theory with asymmetric recombination}

An alternative approach to obtaining a master equation for the ensemble averaged density operator is to apply Redfield theory.\cite{Redfield1965,Goldman2001,Wangsness1953} Here we demonstrate how to do this for the radical pair problem, including the effect of asymmetric recombination. We start by re-writing the Stochastic Liouville equation in the interaction picture. The equation of motion for the interaction picture density operator, $\op{\rho}^\ipi(t,\sX) = e^{-\pL_0 t}\op{\rho}(t,\sX)$, is
\begin{align}
\pdv{t}\op{\rho}^\ipi (t,\sX) = \pL_V^\ipi(t)\op{\rho}^\ipi(t,\sX),\label{sle-ipi-eq}
\end{align}
in which the interaction picture Liouvillian is $\pL_V^\ipi(t) = e^{-\pL_0 t}\pL_V e^{\pL_0 t}$, and $\pL_0$ and $\pL_V$ have the same definitions as above. The solution to this equation is
\begin{align}
\op{\rho}^\ipi (t,\sX) = \mathsf{T}\exp(\int_0^t \pL_V^\ipi(\tau)\dd{\tau})p_0(\sX) \ev{\op{\rho}(0)},
\end{align}
\highlight{where $\mathsf{T}$ is the time-ordering operator.} Taking the ensemble average of this and we find the formal solution for the ensemble averaged density operator,
\begin{align}
\ev{\op{\rho}^\ipi (t)} = \ev{\mathsf{T}\exp(\int_0^t\dd{\tau} \pL_V^\ipi(\tau) ) }\ev{\op{\rho}(0)}.\label{ipi-formal-sol-eqn}
\end{align}
We now make the second order cumulant approximation to the ensemble averaged propagator,\cite{Freed1968} noting that the first order term vanishes because $\ev{\pL_V}=0$,
\begin{align}
\begin{split}
&\ev{\mathsf{T} \exp(\int_0^t \dd{\tau}\pL_V^\ipi(\tau) ) } \\
&\approx \mathsf{T} \exp( \int_0^t\dd{\tau}\int_{0}^\tau\dd{\tau'}\ev{ \pL_V^\ipi(\tau) \pL_V^\ipi(\tau') } ).
\end{split}
\end{align}
Inserting this into Eq. \eqref{ipi-formal-sol-eqn}, differentiating with respect to $t$, and moving back to the Schr\"odinger picture, we find 
\begin{align}
\dv{t}\ev{\op{\rho} (t)} =  -\frac{i}{\hbar} [\op{H}_0,\ev{\op{\rho} (t)}] +\pazocal{R}_\mathrm{RF}(t)\ev{\op{\rho} (t)},\label{avsle2-eq}
\end{align}
in which the time-dependent Redfield superoperator can be written as
\begin{align}
&\pazocal{R}_\mathrm{RF}(t) = \int_0^t \dd{\tau}\int \dd{\sX}{\pL_Ve^{\pL_0(t-\tau)}\pL_Ve^{-\pL_0(t-\tau)}}p_0(\sX) \\
&=-\sum_{j,k}\int_0^t\dd{\tau} g_{jk}(\tau)\pazocal{A}^\dag_je^{\ev{\pL}\tau}\pazocal{A}_ke^{-\ev{\pL}\tau}.\label{rf-superop-eq}
\end{align}
Finally, if the correlation functions, $g_{jk}(\tau)$, decay on a much shorter time-scale than the dynamics of $\ev*{\op{\rho}(t)}$, then we can approximate $\pazocal{R}_\mathrm{RF}(t)$ by replacing $t$ with $\infty$ in the integral limit. 
This gives the standard Markovian and time-homogeneous Redfield master equation,\cite{Goldman2001}
\begin{align}
\dv{t}\ev{\op{\rho} (t)} =  -\frac{i}{\hbar} [\op{H}_0,\ev{\op{\rho} (t)}] - \{\op{K},\ev{\op{\rho}(t)} \} + \pazocal{R}_\mathrm{RF}\ev{\op{\rho} (t)},\label{rf-eq}
\end{align}
where the Redfield relaxation superoperator is defined as $\pazocal{R}_\mathrm{RF}\equiv\lim_{t\to\infty}\pazocal{R}_\mathrm{RF}(t)$. 

It should be noted that this version of Redfield theory explicitly includes the effect of asymmetric radical pair recombination ($k_\sing\neq k_\trip$) on the relaxation processes, although to the best of our knowledge, asymmetric recombination is normally ignored in evaluating this superoperator.\cite{Lukzen2017,Kattnig2016a,Steiner2018} It can be easily verified that our $\pR_{\rm RF}$ reduces to the standard Redfield superoperator in the case of symmetric recombination ($k_\sing=k_\mathrm{T}$). 

\subsection{Comparison of Nakajima-Zwanzig and Redfield theories}

\begin{figure}
	\includegraphics[width=0.45\textwidth]{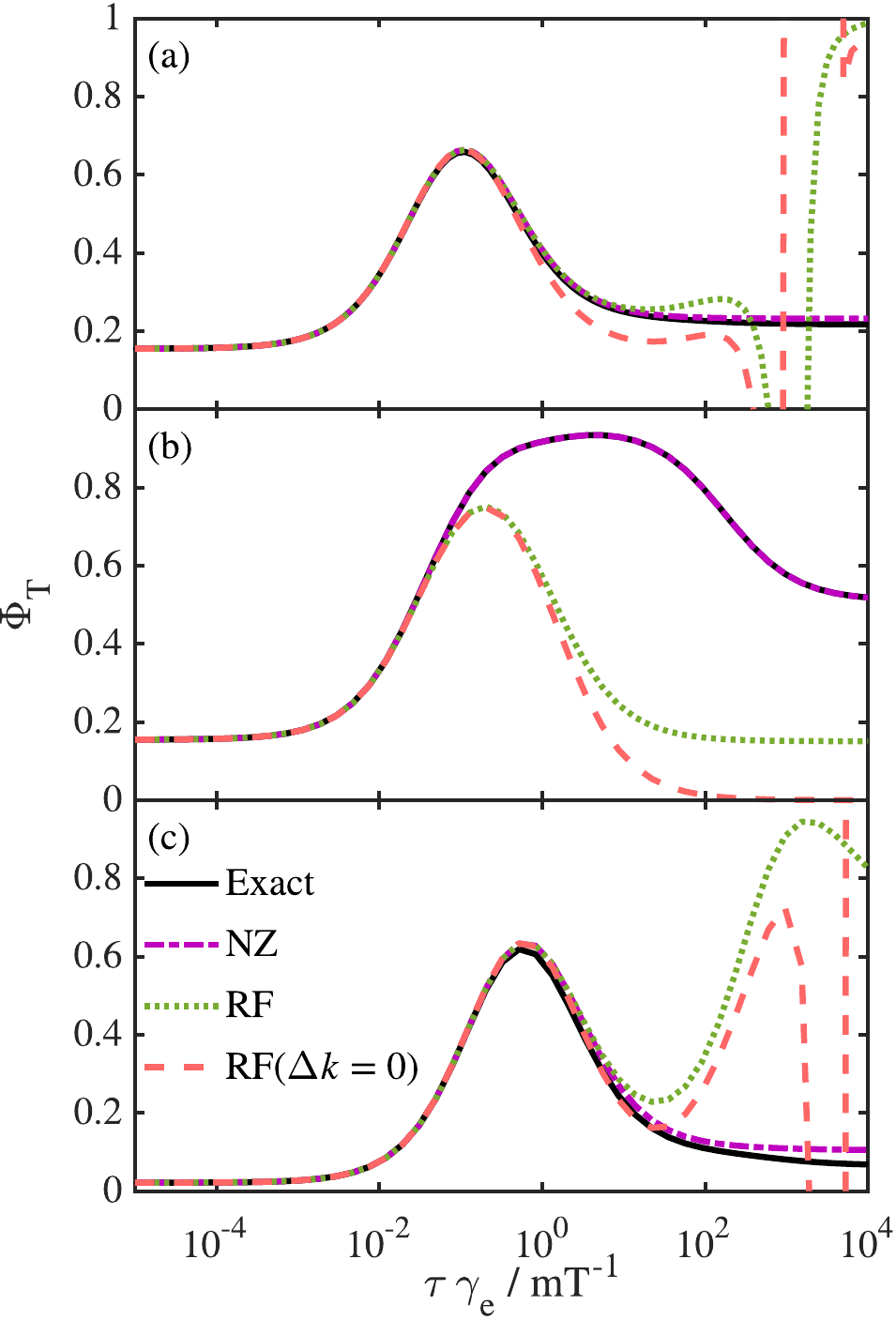}
	\caption{ A comparison of the exact Stochastic Liouville Equation results with the Nakajima-Zwanzig and Redfield approximations to the triplet yield of a one proton radical pair, with three different relaxation mechanisms. In (a) rotational diffusion modulates the anisotropic hyperfine coupling, in (b) the radical pair jumps symmetrically between two sites with different values of the scalar electron spin coupling and in (c) the radical pair asymmetrically jumps between two sites with different values of the isotropic hyperfine coupling. }\label{rf-fig}
\end{figure}

The Nakajima-Zwanzig and Redfield relaxation master equations are both Markovian, time-homogeneous equations for the ensemble-averaged density operator derived from a second order treatment of the fluctuation term in the Liouvillian. However, the two relaxation superoperators are subtly different.{\footnote{{The second order Markovian approximation to Nakajima-Zwanzig theory is sometimes referred to as Redfield theory in the open quantum systems literature. This is subtly different to Bloch-Redfield-Wangsness theory (which we refer to here as Redfield theory), which has exclusively been used in the recent spin dynamics literature.}}} \highlight{Comparing the Redfield relaxation superoperator [Eq. \eqref{rf-superop-eq}] with the Nakajima-Zwanzig superoperator [Eq. \eqref{nz-superop-eq}], we see that the Redfield superoperator contains an additional backwards-time propagator, $e^{-\ev{\pL}\tau}$, which comes from applying the perturbative approximation in the interaction picture.} In the extreme-narrowing limit, when the decay time of the $g_{jk}(t)$ is much shorter than that of $\ev{\op{\rho}(t)}$, such that we can approximate $g_{jk}(t)\approx\tau_{jk}\delta(t)$, both theories give the same relaxation superoperator. This is not the case in the static disorder, or long-correlation time, limit. \textcolor{black}{In this case neither perturbative master equation necessarily preserves positivity of the density operator. The Nakajima-Zwanzig equation however gives time-integrated observables, such as the radical pair lifetime and triplet yield, accurate to second order in the fluctuation for all fluctuation time-scales, which Redfield theory does not.}

We illustrate this difference between the different master equations for a one proton radical pair in Fig. \ref{rf-fig}. We calculate the triplet yield as a function of the correlation time, $\tau$, for three different relaxation mechanisms, exactly solving the Stochastic Liouville Equation and using the perturbative methods described above. In panel (a) the relaxation mechanism is the modulation of an anisotropic hyperfine coupling by isotropic rotational diffusion, with $1/\tau = 6D$ where $D$ is the rotational diffusion constant. In panel (b) the relaxation mechanism is the modulation of the exchange coupling by stochastic jumps between two sites, with $1/\tau = k_{1\to 2}=k_{2\to 1}$. For the results in panel (c) the relaxation mechanism is the modulation of the proton's isotropic hyperfine coupling to radical electron spin 1 by stochastic jumps between two sites, with $1/\tau = k_{1\to 2}=k_{2\to 1}/5$. In each case $k_\sing = k_\trip/100 = 1\ \mu\mathrm{s}^{-1}$. In (a) and (b) the isotropic hyperfine coupling between the electron spin of radical 1 and the proton is $a = 1\ \mathrm{mT}$, the average electron spin coupling is $\ev{J} = 5\ \mathrm{mT}$, $g_1 = g_2 = g_\el$, and $|\vb{B}| = 2\ \mathrm{mT}$. In (a) the full hyperfine coupling tensor is diagonal in the molecular frame with principal components $A_{xx} = A_{yy} = 0.5\ \mathrm{mT}$ and $A_{zz} = 2\ \mathrm{mT}$, in (b) the scalar electron spin coupling constants for sites 1 and 2 are $J_1 = 0\ \mathrm{mT}$ and $J_2 = 10\ \mathrm{mT}$, and in (c) the isotropic hyperfine coupling constants for the two sites are $a_1 = 0\ \mathrm{mT}$ and $a_2 = 2 \ \mathrm{mT}$.

For each example shown in Fig.~\ref{rf-fig}, the Redfield and Nakajima-Zwanzig approximations agree with the exact result in the extreme narrowing limit, but deviate more as the correlation time of the relaxation process increases. The Redfield results (both with and without the inclusion of the asymmetric recombination in the evaluation of $\pR_\mathrm{RF}$) deviate significantly from the exact results when $1/\tau$ becomes smaller than the characteristic frequency of the spin dynamics, which occurs at $\tau\gamma_\mathrm{e} \approx 1\ \mathrm{mT}^{-1}$. However the Nakajima-Zwanzig equation does not break down in this limit, and the second order Nakajima-Zwanzig equation is in fact exact for the symmetric two site model, as demonstrated in panel (b). \highlight{Whilst neither the Nakajima-Zwanzig equation nor the Redfield equation strictly preserves the positivity of the density operator, the fact that the Nakajima-Zwanzig equation is accurate to second order for time-averaged properties\cite{Fay2019} appears to reduce the significance of the positivity problem in this formulation. Redfield theory is only formally accurate to zeroth order in perturbation theory for time-averaged properties, so the positivity problem in this case is more severe.}

The Nakajima-Zwanzig relaxation superoperator is no more complicated to evaluate than the Redfield superoperator, as we shall show in the appendix. Moreover the results in Fig.~1 show that Nakajima-Zwanzig theory does not suffer from the severe positivity problem of Redfield theory in the static disorder limit. We will therefore use Nakajima-Zwanzig theory in the remainder of this paper to describe relaxation processes in larger and more realistic models of radical pair reactions.

\subsection{The Schulten-Wolynes approximation}

Thus far we have demonstrated that the Nakajima-Zwanzig equation provides an accurate description of relaxation processes in radical pairs spanning the full range of fluctuation correlation times. However, solving this equation without any further approximation has a computational effort that increases exponentially with the size of the spin system. This is problematic because real organic radical pairs often have more than 20 hyperfine coupled nuclear spins, which makes solving the Nakajima-Zwanzig equation quite impractical. It is therefore fortunate that, for short-lived radical pairs and radical pairs in which the electron spin interactions are much stronger than the electron-nuclear spin interactions, the full solution of the Nakajima-Zwanzig equation is not necessary, because the Schulten-Wolynes semiclassical approximation provides a sufficiently accurate approximation to the spin dynamics.\cite{Schulten1978}

\begin{figure}
	\includegraphics[width=0.4\textwidth]{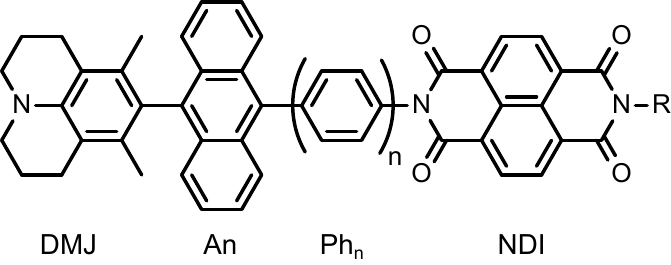}
	\caption{The chemical structure of the DMJ-An-Ph\textsubscript{n}-NDI molecules, where R=n-C\textsubscript{8}H\textsubscript{17}.}\label{rp-struc-fig}
\end{figure}

In the Schulten-Wolynes approximation, the nuclear spin operators in the Hamiltonian are replaced with static classical vectors, $\vb{I}_{i,k}$, of length $\sqrt{I_{i,k}(I_{i,k}+1)}$, each sampled from the surface of a sphere.\footnote{In the original formulation by Schulten and Wolynes, the central limit theorem was used to obtain the distribution of the total nuclear hyperfine field. However in the present formulation we require the orientation of each classical vector to calculate the relaxation superoperators, so we do not make this additional approximation.} The reduced density operator for the electron spin subsystem is approximated as the average over all orientations of the nuclear spin vectors,
\begin{align}
\Tr_\mathrm{nuc}[{\op{\rho}(t,\sX)}]\approx \int {\op{\rho}_\mathrm{SW}(t,\sX,\vb{I})}P(\vb{I})\dd{\vb{I}}.
\end{align}
Here $\vb{I} = (\vb{I}_{1,1},...,\vb{I}_{1,N_1},\vb{I}_{2,1},..,\vb{I}_{2,N_2})$ is the set of nuclear spin vectors, $\int\dd{\vb{I}}P(\vb{I})$ denotes an integral over their orientations, and $\op{\rho}_\mathrm{SW}(t,\vb{I})$ is the Schulten-Wolynes electron spin density operator for a given realisation of the nuclear spins. In practice the integral is evaluated numerically using Monte-Carlo sampling. The Schulten-Wolynes density operator obeys the following equation of motion,
\begin{align}
\begin{split}
\pdv{t} \op{\rho}_\mathrm{SW}(t,\sX,\vb{I}) = &-\frac{i}{\hbar}\left[\op{H}_\mathrm{SW}(\sX,\vb{I}),\op{\rho}_\mathrm{SW}(t,\sX,\vb{I})\right] \\
&- \left\{\op{K},\op{\rho}_\mathrm{SW}(t,\sX,\vb{I})\right\} - \mathsf{D}\,\op{\rho}_\mathrm{SW}(t,\sX,\vb{I}),
\end{split}
\end{align}
where $\op{H}_\mathrm{SW}(\sX,\vb{I})$ is $\op{H}(\sX)$ with the nuclear spin operators replaced by classical vectors ($\op{\vb{I}}_{i,k}\to\vb{I}_{i,k}$). In order to obtain the ensemble dynamics, we can apply perturbative Nakajima-Zwanzig theory directly to this equation of motion, exactly as above. In this case, for each realisation {\bf I} of the nuclear spin vectors, we evolve the ensemble averaged electron spin density operator using the Nakajima-Zwanzig equation,
\begin{align}
\begin{split}
\dv{t} &\ev{\op{\rho}_\mathrm{SW}(t,\vb{I})} = -\frac{i}{\hbar}\left[\op{H}_{0,\mathrm{SW}}(\vb{I}),\ev{\op{\rho}_\mathrm{SW}(t,\vb{I})}\right] \\
&- \left\{\op{K},\ev{\op{\rho}_\mathrm{SW}(t,\vb{I})}\right\} - \pazocal{R}_{\mathrm{NZ,SW}}(\vb{I})\ev{\op{\rho}_\mathrm{SW}(t,\vb{I})},
\end{split}
\end{align}
where $\op{H}_{0,\mathrm{SW}}(\vb{I})$ and $\pazocal{R}_{\mathrm{NZ,SW}}(\vb{I})$ are the Schulten-Wolynes average Hamiltonian and relaxation superoperator for a given {\bf I}. To construct these we simply replace the nuclear spin operators in $\op{H}_0$ and $\pazocal{R}_{\mathrm{NZ}}$ with the corresponding nuclear spin vectors, as in constructing $\op{H}_{\mathrm{SW}}(\sX,\vb{I})$. This Nakajima-Zwanzig/Schulten-Wolynes (NZ/SW) method consistently combines the Schulten-Wolynes semiclassical approximation with the second order relaxation theory provided by the Nakajima-Zwanzig equation, and so provides a way to efficiently model the spin dynamics of radical pairs containing over 20 coupled spins. With this method we can treat relaxation effects rigorously with microscopic models of the relaxation mechanisms, without the need for phenomenological relaxation terms in the master equation. (The accuracy of the method will be demonstrated in a future paper by comparison with an exact, but considerably more computationally expensive, method for modelling the spin dynamics of radical pair reactions including the effects of spin relaxation.\cite{LindoyInPrep}) 

\section{DMJ-NDI radical pairs}

As an example application of the NZ/SW method outlined above, we shall now use it to investigate the intersystem crossing and charge transfer dynamics of photoexcited dimethyljulolidine (DMJ) anthracene (An) para-oligophenylene (Ph\textsubscript{n}) naphthalenediimide (NDI) molecules, the chemical structure of which is shown in Fig.~\ref{rp-struc-fig}. These molecules have been studied experimentally by Scott \textit{et al.},\cite{Scott2009a} and we will use various models for the spin relaxation to interpret their experimental data for the magnetic field effects on the radical pair lifetime and triplet product quantum yields of the molecules with n=1 and n=2 para-phenylene spacers.

\subsection{Photophysics}

\begin{figure}
	\includegraphics[width=0.45\textwidth]{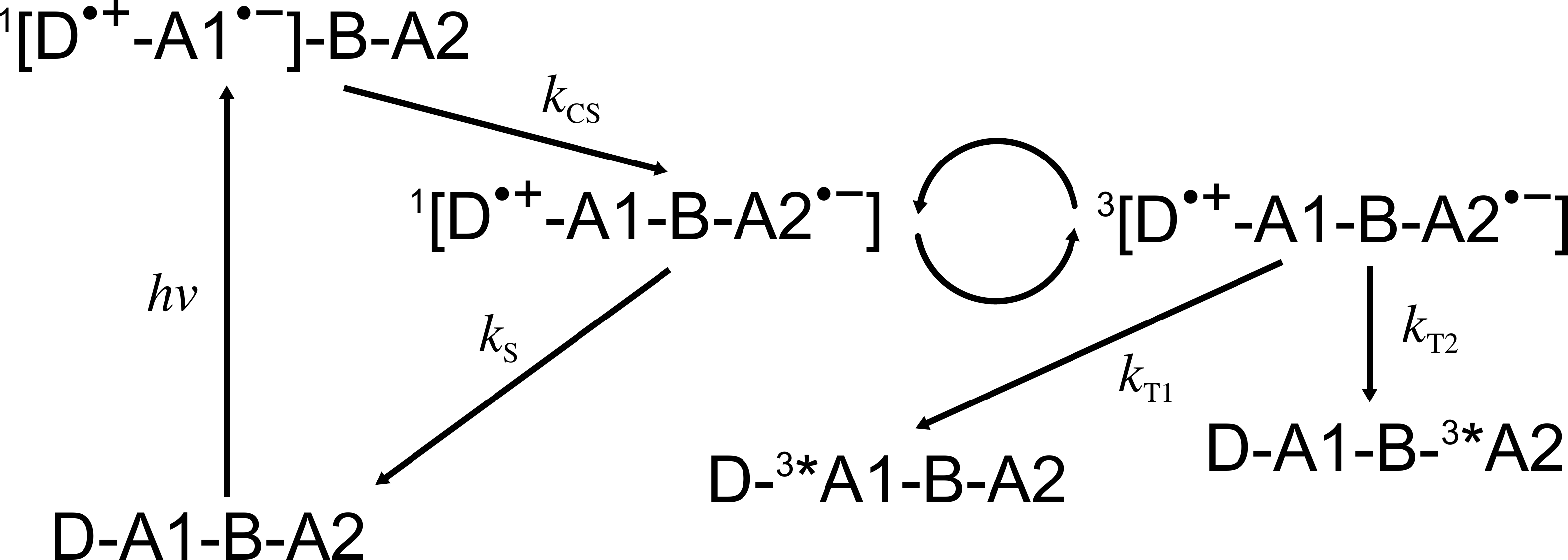}
	\caption{The photophysics of the DMJ-An-Ph\textsubscript{n}-NDI molecule. Here D = DMJ, A1 = An, B = Ph\textsubscript{n} and A2 = NDI. The straight arrows represent incoherent processes and the curved arrows represent the radical pair dynamics. }
\end{figure}

Full details of Scott \textit{et al.}'s experiments can be found in Ref.~\onlinecite{Scott2009a}, where the molecule and its photophysics are fully characterised. Here it suffices to summarise the experiments on the radical pairs, which shall we do with reference to Fig.~3. First the DMJ-An section of the molecule is photoexcited at the edge of its charge transfer band at 416 nm with a laser pulse at 430 nm to form a charge transfer state ${}^1[\text{DMJ}^{\bullet +}\text{-An}^{\bullet -}]$. This then undergoes another charge transfer to form a radical pair state ${}^1[\text{DMJ}^{\bullet +}\text{-An-Ph}_\text{n}\text{-NDI}^{\bullet -}]$ at a rate $k_{\mathrm{CS}}$. The radical pair is formed in the singlet state and undergoes charge recombination to reform the ground state molecule at a rate $k_\sing$, and hyperfine mediated intersystem crossing to form the triplet radical pair state, ${}^3[\text{DMJ}^{\bullet +}\text{-An-Ph}_\text{n}\text{-NDI}^{\bullet -}]$. The triplet radical pair state can then convert back to the singlet radical pair state or undergo a charge transfer to form one of two triplet products. In the triplet products the electronic excitation is localised either on the anthracene, ${}^{3*}\text{An}$, which is formed at a rate $k_{\trip 1}$, or the naphthalenediimide, ${}^{3*}\text{NDI}$, formed at a rate $k_{\trip 2}$. The total triplet recombination rate is then $k_{\trip} = k_{\trip 1}+k_{\trip 2}$. Both the triplet products and the radical pair state are distinguishable spectroscopically, which allows the lifetime of the radical pair state and the triplet product yields to be measured.\cite{Scott2009a} 

\subsection{Magnetic field effects}

Scott {\em et al.}\cite{Scott2009a} measured magnetic field effects on both the lifetime of the radical pair state and the quantum yields of the triplet products relative to the zero field value. Both of these magnetic field effects exhibit a resonance centred at $B=2J$, where $2J$ is the average scalar coupling in the radical pair -- the field at which the singlet and $\trip_+$ triplet states are degenerate in the absence of hyperfine interactions. 

The spin selective charge transfer rates $k_\sing$ and $k_{\trip}$ were estimated experimentally by fitting the kinetic traces of the radical pair state population at different applied field strengths to kinetic models of the radical pair reaction.\cite{Scott2009a} This kinetic modelling approximates the coherent intersystem crossing and incoherent relaxation effects with a simple incoherent model involving field-dependent intersystem crossing rate constants. The approximation provides estimates of recombination rate constants in a straightforward manner, but it ignores the details of the radical pair dynamics. In this work we will apply the NZ/SW method outlined above to compute the lifetimes and relative triplet yields (RTY) of the ${}^{3*}\text{An}$ products of these radical pairs. In doing so we can obtain more reliable estimates of the spin selective recombination rate constants, $k_\sing$ and $k_{\trip}$, as well as determine the relative importance of various relaxation mechanisms and any additional mechanisms by which triplet products can be formed.

\subsection{Relaxation mechanisms}

In our modelling of magnetic field effects on DMJ\textsuperscript{$\bullet+$}-An-Ph\textsubscript{n}-NDI\textsuperscript{$\bullet-$} recombination reactions, we will consider three mechanisms of spin relaxation: rotational diffusion modulating anisotropic spin-spin couplings and the g-tensors, internal motion of the radical pair modulating the scalar electron spin coupling, and conformational changes of the radicals modulating hyperfine coupling between the nuclear and electron spins. To cover all of these cases, the fluctuation Hamiltonian, $\op{V}(\sX)$, is split into rotational motion and internal motion terms. The former depend on the orientation of the radical pair, $\Omega$, and the latter on a set of reduced internal coordinates, $\mathsf{Q}$, which we assume to be uncorrelated with $\Omega$:
\begin{align}
\op{V}(\sX) = \op{V}_\mathrm{rot}(\Omega) + \op{V}_\mathrm{int}(\mathsf{Q}).
\end{align}
In the following we will briefly discuss the rotational and internal motion relaxation mechanisms, and how the theory outlined above can be used to model them.

\subsubsection{Rotational diffusion}

Rotational diffusion of the radical pair leads to a fluctuation in the anisotropic parts of the spin-spin coupling tensors and the anisotropic components of the g-tensors. The overall fluctuation has the form\cite{Worster2016,Nicholas2010}
\begin{align}
\op{V}_\mathrm{rot}(\Omega) = \sum_{m=-2}^2\sum_{m'=-2}^2 \mathfrak{D}_{m',m}^{(2)}(\Omega) \op{Q}^{(2)}_{m,m'},
\end{align}
where $\op{Q}^{(2)}_{m,m'}=\op{Q}_{1,m,m'}^{(2)}+\op{Q}_{2,m,m'}^{(2)}+\op{Q}_{\mathrm{dip},m,m'}^{(2)}$. Here $\op{Q}^{(2)}_{i,m,m'}$ is the anisotropic component of the Hamiltonian of radical $i$,
\begin{align}
\op{Q}^{(2)}_{i,m,m'} = \mu_\mathrm{B} g_{i,m}^{(2)}T_{m'}^{(2)}(\op{\vb{S}}_i,\vb{B}_0)+\sum_{k=1}^{N_i}A_{i,k,m}^{(2)}T_{m'}^{(2)}(\op{\vb{S}}_i,\op{\vb{I}}_{i,k}),
\end{align}
and $\op{Q}^{(2)}_{\mathrm{dip},m,m'}$ is the dipolar electron spin coupling term
\begin{align}
\op{Q}^{(2)}_{\mathrm{dip},m,m'} = D_{m}^{(2)}T_{m'}^{(2)}(\op{\vb{S}}_1,\op{\vb{S}}_2).
\end{align}
In these expressions, $T_{m'}^{(2)}(\vb{u},\vb{v})$ are rank 2 spherical tensor components for a pair of vectors $\vb{u}$ and $\vb{v}$, $A_{i,k,m}^{(2)}$, $g_{i,m}^{(2)}$ and $D_{m}^{(2)}$ are the rank 2 spherical tensor components of the hyperfine coupling tensors, g-tensors, and dipolar coupling tensor, respectively.\cite{Nicholas2010} $\mathfrak{D}_{m',m}^{(2)}(\Omega)$ is the rank 2 Wigner D-matrix element for the molecule in an orientation given by $\Omega$, which is the orientation of the principal axes of the molecule relative to the lab frame.\cite{Nicholas2010}

This approximates the radical pair as a rigid body, and we will further limit our discussion to the case of symmetric top diffusion. In this case the $\mathfrak{D}_{m',m}^{(2)}(\Omega)$ matrix elements have the following correlation functions,\cite{Tarroni1991}
\begin{align}
\ev{\mathfrak{D}_{m,n}^{(2)}(0)^*\mathfrak{D}_{m',n'}^{(2)}(\tau)} = \frac{\delta_{m,m'}\delta_{n,n'}}{5} e^{-(6 D_\perp + (D_\parallel-D_\perp) n^2)|\tau|},
\end{align}
where $D_\perp$ and $D_\parallel$ are the rotational diffusion constants perpendicular and parallel to the molecular symmetry axis.

\subsubsection{Internal motion}

Internal vibrational motion of the radical pair modulates isotropic and anisotropic coupling parameters in the radical pair Hamiltonian. Torsional motions of the radicals and the paraphenylene units modulate the superexchange coupling between radicals, leading to a fluctuation term of the form
\begin{align}
\op{V}_\mathrm{int}(\mathsf{Q}) = -2\Delta J(\mathsf{Q}) \op{\vb{S}}_1\cdot\op{\vb{S}}_2.
\end{align}
\textcolor{black}{We model the fluctuation autocorrelation function for $\Delta J$ as a single exponential decay}
\begin{align}
\ev{\Delta J(0)\Delta J(\tau)} = \sigma_J^2 e^{-|\tau|/\tau_J},
\end{align}
where $\sigma_J^2$ is the mean square fluctuation in $J(t)$ and $\tau_J$ is the correlation time of this fluctuation. This model is a significant simplification of the true molecular motion modulating the scalar coupling of the electrons, but it contains a minimal number of parameters.

\begin{figure}
	\includegraphics[width=0.25\textwidth]{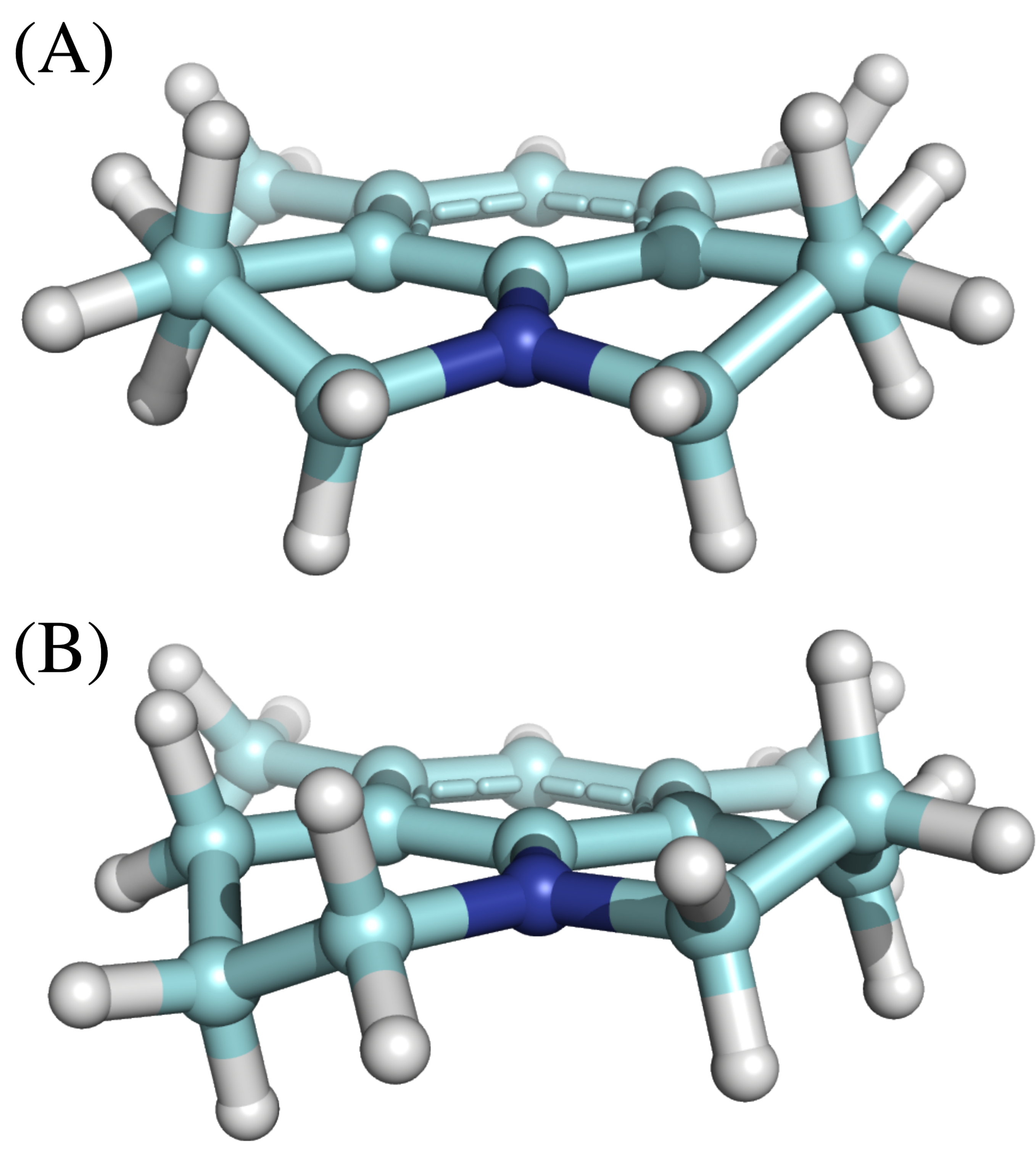}
	\caption{Two of the four conformations of the DMJ radical cation: (A) one of the two \textit{syn} conformations and (B) one of the two \textit{anti} conformations. Geometries calculated using DFT with the B3LYP functional with D3 dispersion correction and the 6-31G(d,p) basis set.}\label{confs-fig}
\end{figure}

The $\text{DMJ}^{\bullet+}$ radical ring system has four stable conformations, two \textit{syn} and two \textit{anti} (see Fig.~\ref{confs-fig}). Each ring can flip, with an \textit{anti} to \textit{syn} rate constant of $k_\mathrm{flip}$, which changes the C--H bonds in the ring hyperconjugated with the nitrogen atom p-orbital. This modulates the proton hyperfine couplings, which gives rise to spin relaxation. We assume that only one ring can flip at a time, so the model includes two rate constants: the \textit{anti}$\to$\textit{syn} flip rate constant is $k_{\mathrm{flip}}$, and the back flipping rate constant is $k_\mathrm{flip}'$. With this model we ignore fluctuations of the anisotropic components of the hyperfine coupling tensor, which are at least 10 times smaller than the fluctuations in the isotropic components.

\subsection{Additional triplet formation mechanisms}

In addition to triplet radical pair recombination, the triplet product can be formed by other mechanisms. This may be required to explain the observed magnetic field effect on the triplet product yield at high magnetic fields, as both we and others have noted previously for the charge recombination along other molecular wires.\cite{Fay2017, Steiner2018, Lukzen2017, Klein2015}

One possibility is that spin-orbit coupling plays a role in the singlet charge recombination.\cite{Miura2010} Some intersystem crossing will then accompany this charge recombination, yielding a triplet product from a singlet radical pair. Assuming this occurs at a rate $k_{\sing\trip 1}$ for ${}^{3*}\text{An}$, the observed triplet yield, $\Phi_{\trip 1,\mathrm{obs}}$ will be related to the quantum yields defined in Eq. \eqref{yields-eq} by
\begin{equation}
\Phi_{\trip 1,\text{obs}} = \frac{k_{\trip 1}}{k_{\trip}}\Phi_\trip + \frac{k_{\sing\trip 1}}{k_{\sing}}\Phi_\sing, 
\end{equation}
where $k_\sing$ and $k_\trip$ are the total recombination rates from the radical pair singlet and triplet states. It follows that the relative triplet yield of the DMJ-$^{3*}$An-Ph$_{\rm n}$-NDI product state will be given by
\begin{equation}
\text{RTY}(B) = {\Phi_{\rm T1,obs}(B)\over \Phi_{\rm T1,obs}(0)} = \frac{\Phi_{\trip}(B)+\Phi_0}{\Phi_{\trip}(0)+\Phi_0},
\end{equation}
where
\begin{equation}
\Phi_0 = {k_{\rm ST1}\over {k_{\rm S}k_{\rm T1}/k_{\rm T}-k_{\rm ST1}}}
\end{equation}
is a field-independent ``background'' contribution to the production of this state.

Another possibility is that in the initial charge separation step, a small fraction of triplet radical pairs are generated.\cite{Maeda2011} In this case the initial condition of the ensemble averaged density operator becomes
\begin{align}
\ev{\hat{\rho}(0)} = \frac{1-\lambda_\trip}{Z}\hat{P}_\sing + \frac{\lambda_\trip}{3Z}\hat{P}_\trip,
\end{align}
where $\lambda_\trip$ is the initial triplet fraction. At high fields the $\trip_+$ and $\trip_-$ triplet states are well separated in energy from the singlet state, so radical pairs created in these states cannot convert to the singlet state and therefore simply recombine to give the triplet product.

The final mechanism for triplet product formation we consider is some additional relaxation not accounted for by our microscopic modelling. Any source of rapidly fluctuating magnetic fields or a mechanism which randomises the electron spin state could give rise to additional relaxation. We allow for such a contribution using the following superoperator,\cite{Kattnig2016}
\begin{align}
\begin{split}
\pL_{\mathrm{rel}}\ev{\hat{\rho}(t)} =\ &-k_\mathrm{rel}\left(\frac{3}{2}-\sum_{\alpha=x,y,z}\op{S}_{1,\alpha}\ev{\rho(t)}\op{S}_{1,\alpha}\right),
\end{split}
\end{align}
in which we choose $\op{S}_{1,\alpha}$ to correspond to the DMJ radical electron spin. This form of the relaxation operator arises from random fields relaxation in the extreme narrowing limit, in which case $k_\mathrm{rel} = 2 \tau_{\mathrm{c}}\ev*{\Delta B^2}\gamma_\mathrm{e}^2$, where $\ev*{\Delta B^2}$ is the mean square fluctuation in the random field and $\tau_{\mathrm{c}}$ is the correlation time of the fluctuation.\cite{Kattnig2016}

\begin{figure*}
	\includegraphics[width = 0.95\textwidth]{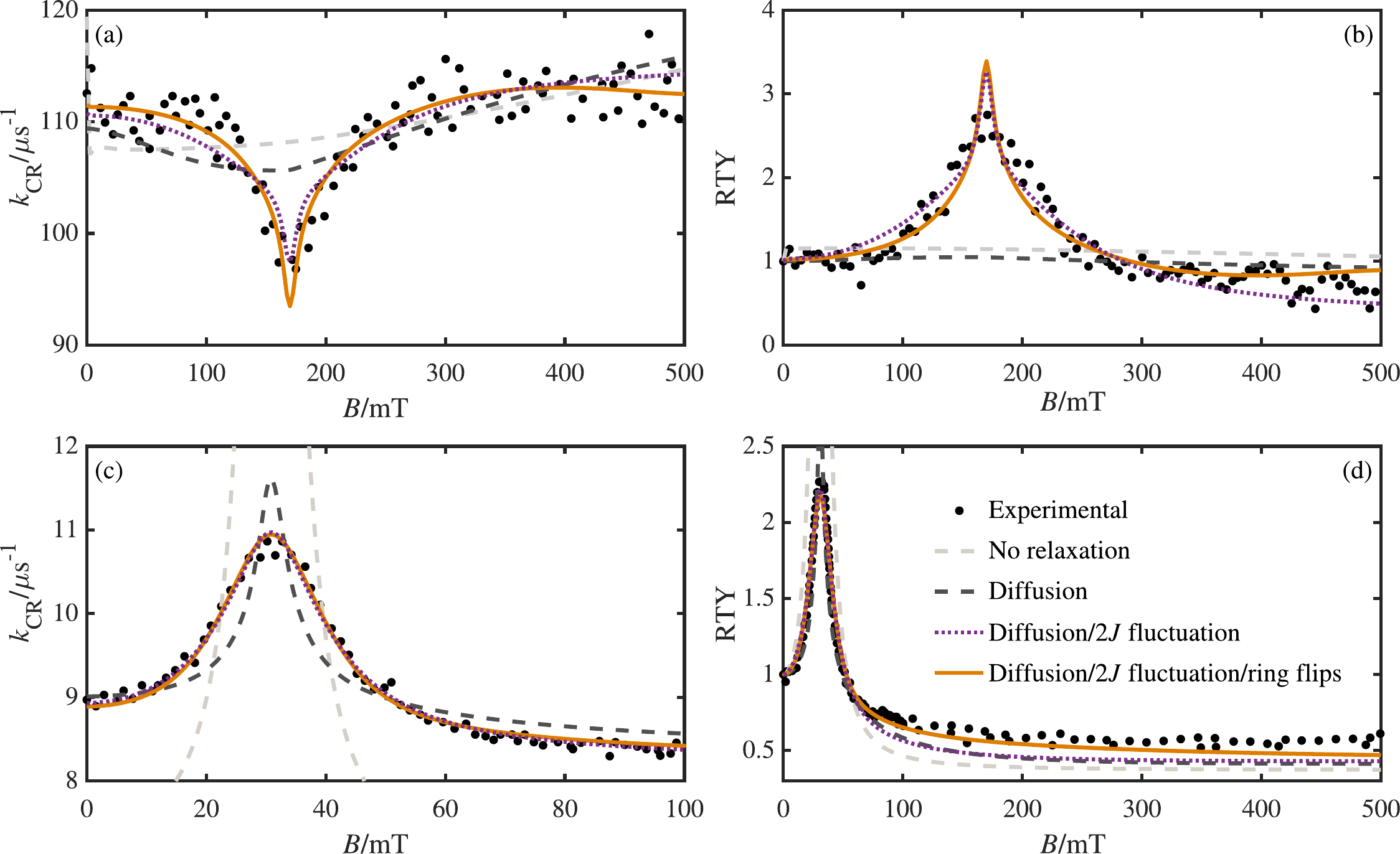}
	\caption{Best fit NZ/SW simulations of the  $k_\mathrm{CR}$ and RTY magnetic field effect data for the n=1 [(a) and (b)] and n=2 [(c) and (d)] molecules with different relaxation mechanisms included. Experimental data are taken from Ref. \onlinecite{Scott2009a}. The simulations used 500 Monte Carlo samples of the static nuclear spin vectors.}\label{relaxation-fig}
\end{figure*}

\subsection{Simulation parameters}

Many of the parameters in our simulations can be determined either from existing experimental data or electronic structure calculations. The average scalar electron spin coupling can be extracted from the peaks in the relative triplet yield and radical pair lifetime curves as a function of applied magnetic field,\cite{Scott2009a} which give $2J=170\text{ mT}$ and $31\text{ mT}$ for the n=1 and n=2 molecules respectively. The hyperfine coupling tensors, $g$-tensors, and dipolar coupling tensors were obtained from available experimental data\cite{Andric2004} and DFT calculations, and the rotational diffusion constants were estimated based on the Stokes-Einstein equation (as outlined in the Supplementary Information). For simulations including the \textit{anti}-\textit{syn} ring flipping of the DMJ radical, the ratio of the flipping rates is given by $k_\mathrm{flip}/k_\mathrm{flip}' = e^{-\Delta G/RT}$. Our results were not found to be strongly dependent on $\Delta G$ so we used the gas phase energy difference of $\Delta G \approx 1.5\text{ kJ mol}^{-1}$, calculated at the DLPNO-CCSD(T)/aug-cc-pVTZ level of theory, as a simple estimate.\cite{Neese2012}

\begin{figure*}
	\includegraphics[width = 0.95\textwidth]{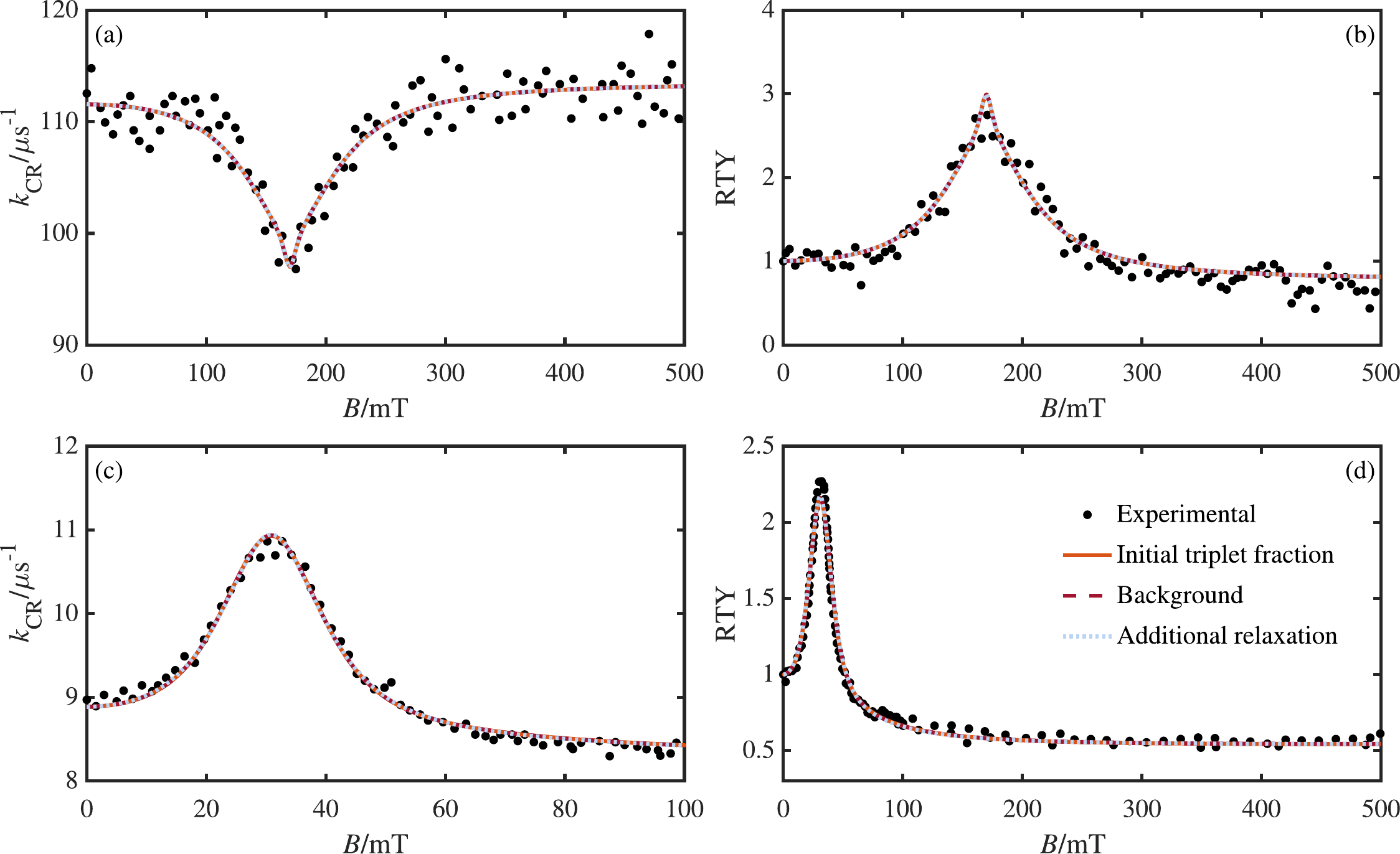}
	\caption{Best fit NZ/SW simulations of the  $k_\mathrm{CR}$ and RTY magnetic field effect data for the n=1 [(a) and (b)] and n=2 [(c) and (d)] molecules with rotational diffusion and $2J$ modulation relaxation mechanisms plus an additional triplet formation mechanism: a background contribution to the triplet yield [Eq.~(40)]; an initial triplet radical pair fraction [Eq.~(42)]; and an additional random field relaxation mechanism [Eq.~(43)]. Experimental data are taken from Ref.~\onlinecite{Scott2009a}. The simulations used 500 Monte Carlo samples of the static nuclear spin vectors.}\label{tripmech-fig}
\end{figure*}

Given that not all parameters are known \textit{a priori}, and that we do not even know which mechanisms of additional triplet formation may operate in these radical pairs, we shall systematically increase the complexity of our model, including more relaxation processes and additional triplet formation mechanisms as we do so. At each stage we need to determine the total spin selective recombination rates, $k_{\sing}$ and $k_{\trip}$, as well as some of the relaxation parameters and additional triplet formation parameters. These are obtained by fitting each model to the experimental charge recombination rate, $k_\mathrm{CR}(B)=1/\tau_\mathrm{RP}(B)$, and relative triplet yield data RTY$(B)$, over a range of magnetic field strengths. For the n=1 molecule, we fit to $k_\mathrm{CR}(B)$ and RTY$(B)$ for magnetic field strengths between 0 mT and 500 mT, and for n=2 we fit to the available $k_\mathrm{CR}(B)$ data between 0 mT and 100 mT and the RTY$(B)$ data between 0 mT and 500 mT.\cite{Scott2009a} The parameters are extracted from the data by minimising an equally weighted sum of the normalised mean square errors (NMSEs) in the fits. This corresponds to minimising the following quantity with respect to the free parameters,
\begin{align}
\begin{split}
\overline{\mathrm{NMSE}} =  &\frac{1}{2N_{k_\mathrm{CR}}} \sum_{i = 1}^{N_{k_\mathrm{CR}}}\frac{(k_\mathrm{CR}^{(\mathrm{exp})}(B_i)-k_\mathrm{CR}^{(\mathrm{calc})}(B_i))^2}{(k_\mathrm{CR,max}^{(\mathrm{exp})}-k_\mathrm{CR,min}^{(\mathrm{exp})})^2} \\
+\frac{1}{2N_\mathrm{RTY}} &\sum_{i = 1}^{N_{\mathrm{RTY}}}\frac{(\mathrm{RTY}^{(\mathrm{exp})}(B_i)-\mathrm{RTY}^{(\mathrm{calc})}(B_i))^2}{(\mathrm{RTY}_\mathrm{max}^{(\mathrm{exp})}-\mathrm{RTY}_\mathrm{min}^{(\mathrm{exp})})^2},
\end{split}
\end{align}
where $k_\mathrm{CR}^{(\mathrm{exp})}(B_i)$ and $\mathrm{RTY}^{(\mathrm{exp})}(B_i)$ are the experimental charge recombination rate and relative triplet yield at field strength $B_i$ and the subscripts $\mathrm{max}$ and $\mathrm{min}$ indicate the maximum and minimum values of these datasets. 


\section{Results}

The best fits to the $k_\mathrm{CR}$ and RTY data for models of the radical pair reaction including different relaxation mechanisms are shown in Fig.~\ref{relaxation-fig}. The fit quality dramatically improves as relaxation by rotational diffusion and $2J$ modulation are included for both molecules (n=1 and n=2); adding relaxation due to ring flips of the DMJ has little effect on the fit quality. However, none of the models including relaxation perfectly describes all of the features of the experimental magnetic field effect data. In particular, the $k_\mathrm{CR}$ $B=2J$ peak shape for the n=1 molecule and the high-field limit of the RTY for the n=2 molecule are not captured well by these models, as can be seen in panels (a) and (d) of the figure.

In Fig.~\ref{tripmech-fig} we display the best fits to the experimental data for simulations including rotational diffusion, $2J$ modulation, and one of the three proposed additional triplet formation mechanisms. Each of the additional triplet formation mechanisms improves the fit to the $B=2J$ peak of the n=1 $k_\mathrm{CR}$ data and the high field limit of the n=2 RTY data, although the quantitative improvement in the $\overline{\mathrm{NMSE}}$ value for n=2 is only small. In these simulations we do not include ring flips since the results in Fig.~\ref{relaxation-fig} suggest they have little influence on the observed magnetic field effects. The final fit quality does not depend on the specific mechanism of additional triplet formation; to graphical accuracy they all produce identical fits. The final fitted $k_{\sing}$, $k_{\trip}$, $2\sigma_J$ and $\tau_J$ parameters, shown in Table \ref{param-tab}, do not vary significantly between the models with the different additional triplet formation mechanisms.

\section{Discussion}

Our fitting of various relaxation models to the experimental magnetic field effect data for the n=1 and n=2 DMJ-NDI molecules suggests that relaxation due to the modulation of anisotropic spin coupling parameters by rotational diffusion and the modulation of $2J$ by internal motions both contribute significantly to the radical pair intersystem crossing dynamics. Our analysis also strongly suggests that some additional triplet formation mechanism is required to explain the experimental magnetic field effects. In the following we will discuss the physical significance of the fitted parameters we have obtained in our spin dynamics models.

\subsection{Rate constants}

Our best fit simulations with additional triplet formation mechanisms give a consistent set of spin-selective radical pair recombination rate constants, varying by only $\sim\!\!\! 5\%$. The values of the rate constants obtained from the simple kinetic model considered in Ref.~\onlinecite{Scott2009a} were $k_\sing = 110\  \mathrm{\mu s}^{-1}$ and $k_\trip = 96\ \mathrm{\mu s}^{-1}$ for n=1 and $k_\sing = 6.8\ \mathrm{\mu s}^{-1}$ and $k_\trip = 15\ \mathrm{\mu s}^{-1}$ for n=2. The largest deviation between our rate constants and those from the kinetic model is in $k_\trip$ for n=1, for which our simulations give a value approximately three times smaller than the kinetic model prediction. In a subsequent paper we shall present the results of fitting the NZ/SW model to exact simulated magnetic field effects.\cite{LindoyInPrep} Based on this, we expect the error from using the NZ/SW method to fit $k_\sing$ and $k_\trip$ to be no more than $\sim\! 5\%$. In any case, given the relatively small differences between our fitted rate constants and those obtained in Ref.~\onlinecite{Scott2009a}, it seems likely as Scott \textit{et al.} concluded that both the singlet and triplet charge transfers proceed via a superexchange mediated tunnelling mechanism.

\subsection{Relaxation mechanisms}

The results in Fig.~\ref{relaxation-fig} show that relaxation arising from rotational diffusion alone is not sufficient to account for the observed magnetic field effect curve peak shapes. The inclusion of modulation of the scalar coupling of the electron spins significantly improves the accuracy of the line shape in both molecules. In particular the widths of the observed peaks are captured well for both n=1 and n=2. Modulation of the scalar coupling gives rise primarily to singlet-triplet dephasing.\cite{Miura2019, Steiner2018, Hoang2018, Miura2010} This increases the decay rate of the singlet-triplet coherences in the radical pair spin dynamics, leading to broadening of the magnetic field effect  peak. Interestingly, however, we have found that a field-independent phenomenological dephasing term in the master equation cannot consistently explain the observed magnetic field effects, especially for n=2. This reflects the fact that the fitted $1/\tau_J$ is $\approx 4\ \mathrm{ns}^{-1}$, whereas an applied field of the order of 500 mT corresponds to a Larmor frequency of the electrons of $\gamma_\mathrm{e} B \approx 88\ \mathrm{ns}^{-1}$. The modulation of $J$ is therefore too slow to be in the extreme narrowing limit and the field dependence of the singlet-triplet dephasing rate needs to be included properly in the calculation, for example by employing \textcolor{black}{an exponential model} for the $J$-modulation as we have done here.

The inclusion of hyperfine tensor modulation by ring flips of the DMJ radical does not significantly improve the fit quality of the simulation to the magnetic field effect data for either molecule. This suggests that the relaxation induced by this hyperfine tensor modulation does not have a large effect on the spin dynamics of the radical pair. Furthermore, because the inclusion of this relaxation mechanism does not help to explain the magnetic field effect data, the $k_\mathrm{flip}$ parameters obtained from the fitting cannot be interpreted as the true physical ring flipping rate constants. \textcolor{black}{We have also} checked that this result is not dependent on our estimate of $k_\mathrm{flip}/k'_\mathrm{flip}$. 

For the $J$ modulation, we find that the size of the fluctuation, $\sigma_J$, scales roughly with average coupling, $\ev{J}$, as is expected. For n=1 the best fit value of $\sigma_J$ is $\approx 1.4 \ev{J}$, and for n=2  it is $\approx 0.2\ev{J}$. The time scales of the fluctuations, $\tau_J$, for the n=1 and n=2 wires differ by a factor of $\approx 70$. This seems a little large to be due to the torsional motions of the molecules modulating the exchange coupling, although the addition of the extra bridging para-phenylene group in the n=2 wire will clearly have some effect on this. \textcolor{black}{It should also be noted that the correlation functions for the true molecular motions that modulate the scalar coupling are likely to be more complex than the simple exponential model we have used here.} 

The difference in $\tau_J$ between the n=1 and n=2 wires could alternatively be explained by electron or hole hopping to a short-lived high-energy state with a larger scalar coupling to the other radical, for example with the bridge in a $\text{Ph}_\text{n}^{\bullet+}$ state or an $\text{An}^{\bullet+}$ state. The lifetime of this excited state could be very different between the n=1 and n=2 molecules, which could explain the large difference in their $\tau_J$s. In particular, an $\text{An}^{\bullet+}$ radical intermediate might account for the slower time-scale and smaller variance of the n=2 $J$ fluctuations than those for n=1, because one would expect the $\text{An}^{\bullet +}{\rm -Ph}_\text{n}\text{-NDI}^{\bullet -}$ radical pair to be less electrostatically stabilised for n=2 than for n=1. 

\highlight{While we cannot conclusively determine which of these mechanisms predominantly controls the $2J$ modulation, we suspect that the large difference in the $\tau_J$s of the two molecules makes electron/hole hopping to intermediate radical pair states the more likely explanation. Further experiments, complimented by an analysis of the type we have performed here, might help to resolve this question. One possibility would be to repeat the MFE measurements in different solvents with similar viscosities but different dielectric constants. If electron/hole hopping controls the $2J$ fluctuations, then the $\tau_J$ values would be expected to have a strong dependence on the solvent polarity, whereas if torsional motions control the $2J$ fluctuations, the solvent polarity would not be expected to have nearly such a large effect on the $\tau_J$s.}

\begin{table*}
	\begin{tabular}{llcccccc}
		\ & Model & $\overline{\mathrm{NMSE}}$ & $k_{\sing,\mathrm{tot}}/\mathrm{\mu s}^{-1}$ & $k_{\trip,\mathrm{tot}}/\mathrm{\mu s}^{-1}$ & $2\sigma_J/\mathrm{mT}$ & $\tau_J/\mathrm{ps}$ & \ \\
		\hline
		n = 1 & No relaxation & $4.3\times 10^{-2}$ & $413.0$ &  $1.178\times10^{-4}$ & {---} & {---} \\
		n = 1 & Rotational diffusion & $4.3\times 10^{-2}$ & $407.5$ & $8.173\times 10^{-3}$ & --- & --- \\
		n = 1 & Rotational diffusion/$2J$ fluctuation & $8.6\times 10^{-3}$ & $117.9$ & $29.01$ & $1120$ & $0.3737$ \\
		n = 1 & Rotational diffusion/$2J$ fluctuation/ring flips & $6.9\times 10^{-3}$ & $121.3$ & $2.072$ & $399.9$ & $21.30$ & $k_\mathrm{flip}=2.558\ \mathrm{ns}^{-1}$\\
		n = 1 & Rotational diffusion/$2J$ fluctuation/additional relaxation & $5.8\times 10^{-3}$ & $118.0$ & $31.75$ & $234.5$ & $3.728$ & $k_\mathrm{rel} = 3.590\ \mathrm{\mu s}^{-1}$ \\
		n = 1 & Rotational diffusion/$2J$ fluctuation/initial triplet fraction & $5.8\times 10^{-3}$ & $118.0$ & $31.75$ & $235.4$ & $3.724$ & $\lambda_\trip = 0.02214$ \\
		n = 1 & Rotational diffusion/$2J$ fluctuation/background & $5.8\times 10^{-3}$ & $114.6$ & $37.50$ & $235.0$ &  $3.739$ &  $\Phi_0 = 0.02462$ \\
		\hline
		n = 2 & No relaxation & $0.4$ & $6.536$ &  $296.8$ & {---} & {---} \\
		n = 2 & Rotational diffusion & $1.1\times 10^{-2}$ & $8.009$ & $16.05$ & --- & --- \\
		n = 2 & Rotational diffusion/$2J$ fluctuation & $2.0\times 10^{-3}$ & $7.746$ & $14.34$ & $7.493$ & $202.8$ \\
		n = 2 & Rotational diffusion/$2J$ fluctuation/ring flips & $1.2\times 10^{-3}$ & $7.660$ & $14.53$ & $6.580$ & $349.3$ &  $k_\mathrm{flip}=26.04\ \mathrm{ns}^{-1}$\\
		n = 2 & Rotational diffusion/$2J$ fluctuation/additional relaxation & $1.05\times 10^{-3}$ & $7.134$ & $13.82$ & $6.159$ & $350.1$ & $k_\mathrm{rel} = 0.6973\ \mathrm{\mu s}^{-1}$ \\
		n = 2 & Rotational diffusion/$2J$ fluctuation/initial triplet fraction & $1.05\times 10^{-3}$ & $7.651$ & $14.20$ & $5.724$ & $411.0$ & $\lambda_\trip = 0.06248$ \\ 
		n = 2 & Rotational diffusion/$2J$ fluctuation/background & $1.05\times 10^{-3}$ & $7.883$ & $14.19$ & $5.823$ &  $395.1$ &  $\Phi_0 = 0.06804$ 
	\end{tabular}
	\caption{Fitted parameters for the different models of DMJ-NDI radical pair spin dynamics, with the corresponding normalised mean square errors as defined in Eq.~(44).}\label{param-tab}
\end{table*}

\subsection{Triplet formation mechanisms}

The quality of the fit of our simulations to the experimental data improves significantly when an additional triplet formation mechanism is included in the simulations (see Fig.~\ref{tripmech-fig}). Interestingly, the improvement is independent of the exact mechanism of triplet formation, and the fitted $k_{\sing}$, $k_{\trip}$, $\sigma_J$ and $\tau_J$ parameters also agree well between the different triplet formation models. The fit quality alone cannot therefore be used to infer which additional triplet formation mechanism plays a role in these radical pairs, but we can at least speculate about which we feel to be the most likely. 

Firstly, it should be noted that the relative triplet yields measured experimentally for the ${}^{3*}\text{An}$ and ${}^{3*}\text{NDI}$ products are very similar.\cite{Scott2009a} Given that the fitted background triplet yields are small, $k_{\sing\trip 1}\ll k_{\sing}$, and therefore $\Phi_0\approx k_{\sing \trip 1}k_{\trip} / k_{\trip 1} k_{\sing}$. If the singlet radical pairs were undergoing spin-orbit coupled charge recombination to give both triplet products, then $k_{\sing\trip 1}$ and $k_{\sing\trip 2}$ would need to satisfy $k_{\sing \trip 1} / k_{\trip 1} \approx k_{\sing \trip 2} / k_{\trip 2}$. In the non-adiabatic limit, each of these ratios is proportional to a ratio of squared electron transfer Hamiltonian matrix elements, so $|\mel*{{}^\sing\mathrm{RP}}{\op{H}}{\mathrm{\trip_1}}|^2/|\mel*{{}^\trip\mathrm{RP}}{\op{H}}{\mathrm{\trip_1}}|^2\approx|\mel*{{}^\sing\mathrm{RP}}{\op{H}}{\mathrm{\trip_2}}|^2/|\mel*{{}^\trip\mathrm{RP}}{\op{H}}{\mathrm{\trip_2}}|^2$. Given that the spin-orbit coupled charge transfer matrix elements, $\mel*{{}^\sing\mathrm{RP}}{\op{H}}{\mathrm{\trip_{i}}}$, depend strongly on the relative orientation of the orbitals involved in the electron transfer,\cite{Dance2006} and in a different way to the spin-conserving matrix elements, it seems unlikely that this condition would be satisfied, and therefore this mechanism probably does not operate in these molecules under these experimental conditions.

We have also considered the  inclusion of an additional relaxation mechanism in the extreme narrowing limit, with a fluctuation correlation time much shorter than the spin dynamics time scale. This produces a field-independent relaxation  which has much the same effect as including a background contribution to the triplet yield, and it can also be used to fit the experimental data (see Fig.~6). However we find that the phenomenological relaxation rates needed to fit the n=1 and n=2 data differ by more than a factor of 6 (see Table~I). The length of the molecule only changes by $\sim\!\! 10\%$ and it is hard to imagine that such a small change in the dimensions of the molecule would produce such a large change in the relaxation rate. For example the spin-rotation interaction would most likely not display this degree of dependence on the molecular geometry. Dipolar interactions with nuclear spins in the solvent (toluene) could give rise to rapidly fluctuating fields at the radicals, but again it is not immediately clear why there would be such a large difference between in the n=1 and n=2 molecules. 

This leaves initial triplet radical pair formation as the remaining possibility for the additional triplet product formation. We have shown that an initial fraction of triplet radical pairs could also explain the experimental magnetic field effect data, with similar initial triplet fractions in both the n=1 and n=2 cases ($\sim\!\! 2\%$ and $\sim\!\! 6\%$, respectively -- see Table~I). These initial triplet fractions are relatively small and are in line with those measured in other organic linked radical pair systems.\cite{Maeda2011} The initial triplet fraction could arise from spin-orbit coupled charge transfer (SOCT) between the initial photoexcited ${}^1$[DMJ${}^{\bullet+}$-An${}^{\bullet -}$]-Ph\textsubscript{n}-NDI molecule and the triplet radical pair state. The SOCT rate, $k_\mathrm{CS,T}$, would be expected to depend exponentially on the radical separation, in the same way as the spin-conserving charge separation rate constant, $k_\mathrm{CS}$. As a result, we would expect $\lambda_\trip = k_\mathrm{CS,T}/(k_\mathrm{CS}+k_\mathrm{CS,T})$ to be approximately the same in both the n=1 and n=2 molecules, which is consistent with what we find in our simulations. The small difference between $\lambda_\trip$ for $n=1$ and $n=2$ might be explained by the fact that the SOCT matrix element depends strongly on the relative orientation of the orbitals involved in the electron transfer, since the angle between the DMJ and NDI groups changes by about $60^\circ$ between the two molecules.

Overall, then, our simulations seem to suggest that additional intersystem crossing, most likely accompanying the initial radical pair formation step, may give rise to additional triplet product formation in these radical pairs. The available experimental data is however insufficient to establish this conclusively because the various different triplet formation mechanisms we have considered give rise to identical RTY curves. It may also be that all of these mechanisms operate to some small degree simultaneously to produce the observed magnetic field effects.

\section{Concluding remarks}

In this paper we have demonstrated how to consistently combine the Nakajima-Zwanzig theory of relaxation with the Schulten-Wolynes approximation to the electron spin dynamics of a radical pair, including the interplay of asymmetric recombination rates and electron spin relaxation. We have also used this NZ/SW method to study the spin relaxation in covalently linked DMJ-NDI radical pairs. Combining it with simple models of the internal molecular motion, we have investigated the role of different relaxation mechanisms in these radical pair reactions. In as far as possible we have accounted for relaxation processes with microscopic models, rather than phenomenologically. This has proven to be important, for example in describing the singlet-triplet dephasing process in these molecules. This dephasing is not in the extreme-narrowing limit and a simple, field-independent dephasing rate would fail to describe the magnetic field effects that have been observed experimentally. 

We have found that modulation of the scalar coupling between the electron spins plays a significant role in the observed magnetic field effects on these radical pairs. However this modulation alone cannot fully explain the magnetic field dependence of the relative triplet yields and radical pair lifetimes. Another possible relaxation mechanism, ring inversions of the DMJ${}^{\bullet+}$ radical modulating the hyperfine coupling constants in this radical, was found to not play any significant role in the electron spin dynamics. Alternative additional triplet formation mechanisms, involving incoherent intersystem crossing in either the charge separation or charge recombination processes, or in the radical pair state, have also been considered. Each of these possibilities was found to improve the fit to the experimental data equally well, but based on physical considerations, a small initial triplet radical pair fraction seems to provide the most plausible explanation for the additional triplet product formation. 

We hope the method we have developed here will prove useful in the interpretation of magnetic field effects on other radical pair reactions. It is certainly more powerful than the use of simple kinetic models for the radical pair dynamics and phenomenological treatments of relaxation, and yet it is still relatively inexpensive to implement. Most of the parameters needed in the modelling can be obtained from DFT calculations, or EPR measurements of the constituent radicals, and the remaining parameters can be fit to experimental magnetic field effect data in a matter of minutes on a desktop computer. The NZ/SW method described here can also be straightforwardly applied to study the EPR spectra of radical pairs, and to describe other coupled spin systems such as radical triads.\cite{Rugg2017,Horwitz2017a,Sampson2019,Kattnig2017,Kattnig2017a} 

In a subsequent paper,\cite{LindoyInPrep} we shall describe a complementary method for treating electron spin relaxation in radical pairs, based on solving a Stochastic Schr\"odinger Equation in the full radical pair spin Hilbert space, with randomly sampled initial nuclear spin coherent states.\cite{Lewis2016} This method is numerically exact, and it can be used to check the accuracy of the NZ/SW approximation in a \lq\lq single shot" calculation once the NZ/SW method has been used to extract appropriate parameters (such as the singlet and triplet recombination rates and the relaxation parameters in Table I) from experimental data.

\section*{Supplementary Information}

The Supplementary Information contains all parameters, either experimental or obtained from DFT calculations, used in the radical pair models presented in the main text.

\begin{acknowledgements}	
	Thomas Fay is supported by a Clarendon Scholarship from Oxford University, an E.A. Haigh Scholarship from Corpus Christi College, Oxford, and by the EPRSC Centre for Doctoral Training in Theory and Modelling in the Chemical Sciences, EPSRC Grant No. EP/L015722/1. Lachlan Lindoy is supported by a Perkin Research Studentship from Magdalen College, Oxford, an Eleanor Sophia Wood Postgraduate Research Travelling Scholarship from the University of Sydney, and by a James Fairfax Oxford Australia Scholarship.	
\end{acknowledgements}
\vspace{-12pt}
\appendix

\section{Evaluating the relaxation superoperators}

In this appendix we outline how the Redfield superoperator with asymmetric recombination and the Nakajima-Zwanzig relaxation superoperator can be evaluated. 
We start by choosing a basis, writing $\ev{\op{\rho}(t)}$ as a vector in this basis, $\boldsymbol{\rho}(t)$, and writing superoperators as matrices. A general superoperator acting on $\ev{\hat{\rho}(t)}$ of the form $\op{A}\ev{\op{\rho}(t)}\op{B}$ can be written as a matrix-vector operation as
\begin{align}
\op{A}\ev{\op{\rho}(t)}\op{B} \to (\vb{A\otimes B}^\mathsf{T})\,\boldsymbol{\rho}(t),
\end{align}
where $\vb{A}$ and $\vb{B}$ are matrix representations of $\op{A}$ anf $\op{B}$ in the chosen basis. For example, the Nakajima-Zwanzig relaxation superoperator matrix, $\mathbfcal{R}_\mathrm{NZ}$, can be written as
\begin{align}
\mathbfcal{R}_\mathrm{NZ} = -\sum_{j,k}\int_0^\infty \!\!\!\!\dd{\tau} g_{jk}(\tau) \mathbfcal{A}^\dag_j e^{\ev{\mathbfcal{L}}\tau}\mathbfcal{A}_k.
\end{align}
We can write $\ev{\mathbfcal{L}}$ in terms of a diagonal matrix of its eigenvalues $i\boldsymbol{\lambda}$ and a matrix of eigenvectors $\mathbfcal{U}$ as
\begin{align}
\ev{\mathbfcal{L}} = i\mathbfcal{U}\boldsymbol{\lambda}\mathbfcal{U}^{-1},
\end{align}
in terms of which the NZ relaxation matrix is
\begin{align}
\mathbfcal{R}_\mathrm{NZ} &= -\sum_{j,k} \mathbfcal{A}^\dag_j\mathbfcal{U}\int_0^\infty \!\!\!\!\dd{\tau} g_{jk}(\tau) e^{i\boldsymbol{\lambda}\tau}\mathbfcal{U}^{-1}\mathbfcal{A}_k \\
&=-\sum_{j,k} \mathbfcal{A}^\dag_j\mathbfcal{U}J_{jk}(\boldsymbol{\lambda}) \mathbfcal{U}^{-1}\mathbfcal{A}_k.
\end{align}
Here we have defined the spectral density $J_{jk}(z) = \int_0^\infty g_{jk}(\tau)e^{iz\tau}\dd{\tau}$. The eigenvalues and eigenvectors can be found straightforwardly for an average Liouvillian matrix of the form 
\begin{align}
\ev{\mathbfcal{L}} = -i \boldsymbol{\Omega}\otimes \vb{1} + i \vb{1}\otimes\boldsymbol{\Omega}^*,
\end{align}
which arises in the case of a radical pair with the asymmetric recombination treated using the Haberkorn superoperator. (In this case $\boldsymbol{\Omega}=\frac{1}{\hbar}\vb{H}_0-i\vb{K}$, where $\vb{H}_0$ is the matrix representation of $\op{H}_0$ and $\vb{K}$ is the matrix representation of $\op{K}$.) For a Liouvillian of this form, the eigenvalue matrix is
\begin{align}
i\boldsymbol{\lambda} = -i \boldsymbol{\omega}\otimes \vb{1} +i \vb{1}\otimes \boldsymbol{\omega}^*
\end{align}
where $\boldsymbol{\omega}$ is the diagonal matrix of eigenvalues of $\boldsymbol{\Omega}$ and the eigenvector matrix is
\begin{align}
\mathbfcal{U} = \vb{U}\otimes \vb{U}^*,
\end{align}
in which $\vb{U}$ is the matrix of eigenvectors of $\boldsymbol{\Omega}$: $\boldsymbol{\Omega}\vb{U} = \vb{U}\boldsymbol{\omega}$.

We can use the same techniques to evaluate the Redfield relaxation matrix including asymmetric recombination. In this case the relaxation matrix is
\begin{align}
{\mathbfcal{R}}_\mathrm{RF} &= - \sum_{jk}{\mathbfcal{A}}_j^\dag {\mathbfcal{U}}({\mathbfcal{J}}_{jk}\circ({\mathbfcal{U}}^{-1}{\mathbfcal{A}}_k{\mathbfcal{U}})){\mathbfcal{U}}^{-1},
\end{align}
in which $\circ$ denotes element-wise product ($[\vb{X}\circ\vb{Y}]_{ab}=[\vb{X}]_{ab}[\vb{Y}]_{ab}$), and ${\mathbfcal{J}}_{jk}$ is a matrix of spectral densities associated with $g_{jk}(\tau)$ evaluated at the eigenvalues, $i\lambda_a$, of $\ev{{\mathbfcal{L}}}$, 
\begin{align}
[{\mathbfcal{J}}_{jk}]_{ab} = J_{jk}(\lambda_b-\lambda_a).
\end{align}

\bibliography{bibliography}

\end{document}


\title{Supplementary Information to ``Spin relaxation in radical pair reactions I: An efficient semiclassical method''}
\author{Thomas P. Fay}
\affiliation{Department of Chemistry, University of Oxford, Physical and Theoretical Chemistry Laboratory, South Parks Road, Oxford, OX1 3QZ, UK}
\author{Lachlan P. Lindoy}
\affiliation{Department of Chemistry, University of Oxford, Physical and Theoretical Chemistry Laboratory, South Parks Road, Oxford, OX1 3QZ, UK}
\author{David E. Manolopoulos}
\affiliation{Department of Chemistry, University of Oxford, Physical and Theoretical Chemistry Laboratory, South Parks Road, Oxford, OX1 3QZ, UK}

\begin{abstract}
	In this supplementary information we describe our model for the DMJ${}^{\bullet+}$--A$\text{n}$--P$\text{h}_\text{n}$--NDI${}^{\bullet-}$ radical ion pair, including values for the hyperfine coupling tensors, $g$ tensors, dipolar coupling tensors and the rotational diffusion parameters used in our simulations.
\end{abstract}

\maketitle

\section{DMJ${}^{\bullet+}$--A$\text{n}$--P$\text{h}_\text{n}$--NDI${}^{\bullet-}$ Model Parameters}

\subsection{Hyperfine coupling tensors}

\begin{figure}[h]
	\includegraphics[width=0.35\textwidth]{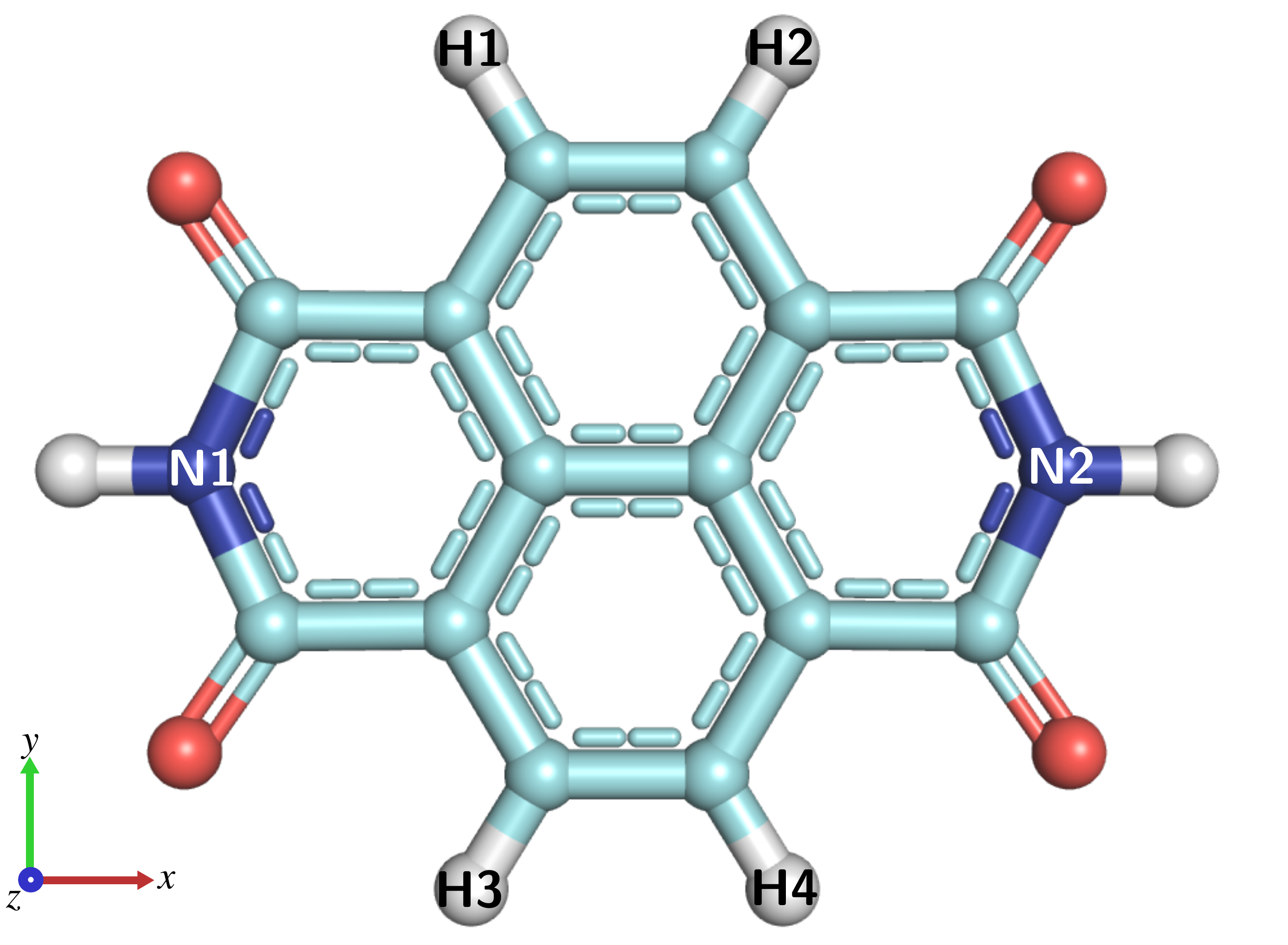}
	\caption{The NDI radical with labelled hyperfine coupled nuclei. The capping hydrogens are not included.}\label{ndi-struc-fig}
\end{figure}

\begin{figure}[h]
	\includegraphics[width=0.65\textwidth]{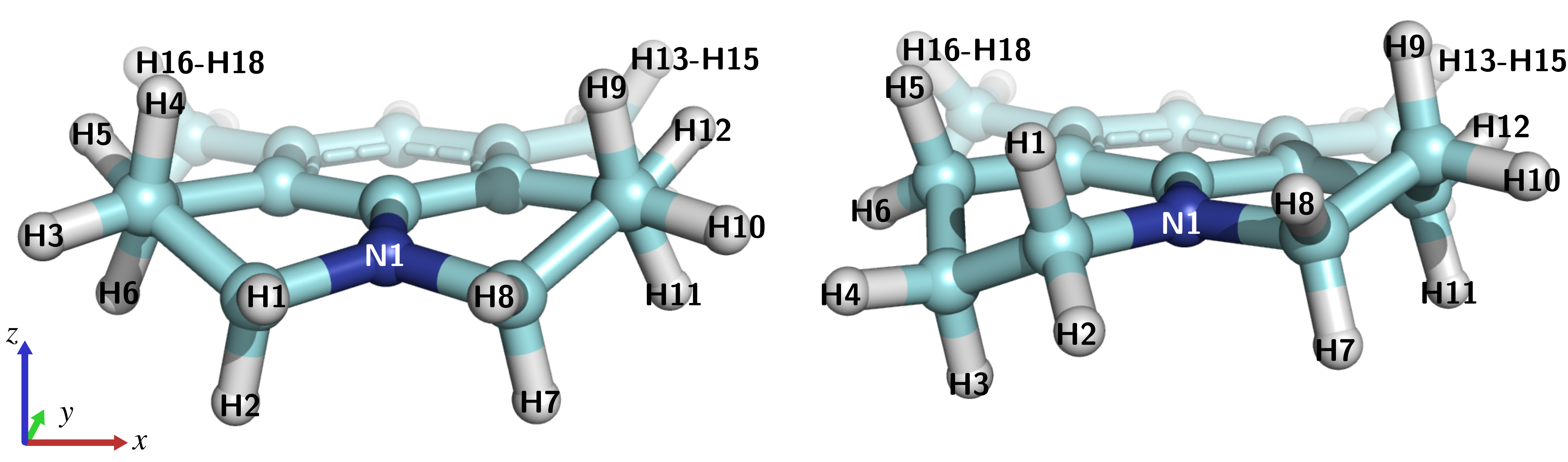}
	\caption{Syn (left) and anti (right) conformations of the DMJ radical with hyperfine coupled nuclei labelled. The capping hydrogen is not included in our simulations.}\label{dmj-struc-fig}
\end{figure}

Isotropic components of the hyperfine coupling tensors, $a_\text{iso} = (A_{xx}+A_{yy}+A_{zz})/3$, for the NDI${}^{\bullet -}$ radical are taken from experimental results\cite{Andric2004} for the \textit{N,N}-dipentyl radical anion. The signs of the isotropic components and the anisotropic components are obtained from DFT. This is done by calculating the radical groundstate geometry, capped with two hydrogens (see Fig. \ref{ndi-struc-fig}), optimized with DFT using the B3LYP functional and 6-31G(d,p) basis set with the D3 dispersion correction, as implemented in Gaussian09, and then calculating the hyperfine coupling tensors with the B3LYP functional and the EPR-III basis set for this optimised geometry. For the NDI${}^{\bullet -}$ radical the isotropic hyperfine coupling constants are found to agree with the experimental values to within $< 10\%$. The hyperfine coupling tensor components are given in Table \ref{ndi-tab}. Isotropic and anisotropic components for the syn and anti conformations of the DMJ${}^{\bullet+}$ radical, capped with hydrogen (see Fig. \ref{dmj-struc-fig}), are calculated in the same way as the anisotropic values for the NDI${}^{\bullet -}$ radical. We additionally average the two sets of three methyl hydrogens, H13--H15 and H16--H18, in both conformations, under the assumption that they rapidly interconvert at 295K. These are given for the syn and anti conformations in Tables \ref{dmj-tab-1} and \ref{dmj-tab-2}. The hyperfine coupling tensors for the other two syn and anti conformations are obtained by inverting the molecular $z$ axis and permuting the labels appropriately. For models which do not include ring flipping of the DMJ, hyperfine tensors for the lowest energy anti conformation is used, but the fit quality and fitted parameters are not found to be sensitive to which of the syn, anti or thermally averaged hyperfine coupling tensors are used for the DMJ${}^{\bullet+}$ radical.

\begin{table}
	\begin{tabular}{clllllll}
		Nucleus & $a_\text{iso}/$mT & $A_{xx}^\text{aniso}$/mT & $A_{yy}^\text{aniso}$/mT & $A_{zz}^\text{aniso}$/mT & $A_{xy}^\text{aniso}$/mT & $A_{xz}^\text{aniso}$/mT & $A_{yz}^\text{aniso}$/mT  \\
		\hline
		H1 & -0.1927 & 0.011586 & 0.032114 & -0.043700 & -0.101834 & -0.000008 & 0.000014 \\
		H2 & -0.1927 & 0.011586 & 0.032114 & -0.043700 & 0.101834 & 0.000014 & 0.000008 \\
		H3 & -0.1927 & 0.011586 & 0.032114 & -0.043700 & 0.101834 & 0.000014 & 0.000008 \\
		H4 & -0.1927 & 0.011586 & 0.032114 & -0.043700 & -0.101834 & -0.000008 & 0.000014 \\
		N1 & -0.0963 & 0.035200 & 0.034000 & -0.069200 & 0.000000 & 0.000000 & 0.000010 \\
		N2 & -0.0963 & 0.035200 & 0.034000 & -0.069200 & 0.000000 & 0.000000 & 0.000010
	\end{tabular}
	\caption{Hyperfine coupling parameters for the NDI${}^{\bullet -}$ radical. Isotropic components, $a_\text{iso} = (A_{xx}+A_{yy}+A_{zz})/3$, are experimental values for the free radical and anisotropic components $A_{\alpha\beta}^\text{aniso} = A_{\alpha\beta}-\delta_{\alpha\beta}a_\text{iso}$ in the isolated radical principal axis frame are obtained from DFT calculations as described in the text.}\label{ndi-tab}
\end{table}

\begin{table}
	\begin{tabular}{clllllll}
		Nucleus & $a_\text{iso}/$mT & $A_{xx}^\text{aniso}$/mT & $A_{yy}^\text{aniso}$/mT & $A_{zz}^\text{aniso}$/mT & $A_{xy}^\text{aniso}$/mT & $A_{xz}^\text{aniso}$/mT & $A_{yz}^\text{aniso}$/mT  \\
		\hline
		H1 & 0.486756 & -0.031726 & 0.144387 & -0.112661 & 0.130562 & -0.015262 & -0.006179 \\ 
		H2 & 2.328949 & 0.029447 & -0.037064 & 0.007617 & 0.096496 & 0.115391 & 0.081504 \\ 
		H3 & -0.061351 & 0.062467 & -0.012795 & -0.049672 & 0.044979 & -0.014692 & 0.001259 \\ 
		H4 & -0.060764 & 0.038959 & -0.043062 & 0.004103 & 0.026643 & -0.086679 & -0.028602 \\ 
		H5 & 0.375611 & 0.092136 & -0.037949 & -0.054187 & -0.025343 & -0.038948 & 0.004200 \\ 
		H6 & 1.153098 & 0.082424 & -0.057527 & -0.024898 & 0.000111 & 0.067331 & -0.006107 \\ 
		H7 & 2.329776 & 0.029272 & -0.036836 & 0.007564 & -0.096634 & -0.115280 & 0.081706 \\ 
		H8 & 0.488286 & -0.032014 & 0.144717 & -0.112703 & -0.130394 & 0.015319 & -0.006342 \\ 
		H9 & -0.060699 & 0.038912 & -0.043011 & 0.004099 & -0.026721 & 0.086652 & -0.028682 \\ 
		H10 & -0.061503 & 0.062395 & -0.012722 & -0.049673 & -0.045031 & 0.014720 & 0.001226 \\ 
		H11 & 1.152785 & 0.082408 & -0.057522 & -0.024886 & -0.000270 & -0.067352 & -0.006030 \\ 
		H12 & 0.375533 & 0.092205 & -0.038003 & -0.054202 & 0.025209 & 0.038923 & 0.004147 \\ 
		H13 & -0.158930 & 0.035962 & 0.000470 & -0.036432 & 0.037651 & 0.000522 & 0.000529 \\ 
		H14 & -0.158930 & 0.035962 & 0.000470 & -0.036432 & 0.037651 & 0.000522 & 0.000529 \\ 
		H15 & -0.158930 & 0.035962 & 0.000470 & -0.036432 & 0.037651 & 0.000522 & 0.000529 \\ 
		H16 & -0.158931 & 0.035888 & 0.000544 & -0.036432 & -0.037686 & -0.000526 & 0.000533 \\ 
		H17 & -0.158931 & 0.035888 & 0.000544 & -0.036432 & -0.037686 & -0.000526 & 0.000533 \\ 
		H18 & -0.158931 & 0.035888 & 0.000544 & -0.036432 & -0.037686 & -0.000526 & 0.000533 \\ 
		N1 & 0.835749 & -0.780100 & -0.737667 & 1.517767 & 0.000068 & -0.000688 & -0.330857 
	\end{tabular}
	\caption{Hyperfine coupling parameters for the DMJ${}^{\bullet +}$ radical in the syn conformation, as labelled above. Isotropic components, $a_\text{iso} = (A_{xx}+A_{yy}+A_{zz})/3$, and anisotropic components, $A_{\alpha\beta}^\text{aniso} = A_{\alpha\beta}-\delta_{\alpha\beta}a_\text{iso}$, in the isolated radical principal axis frame are obtained from DFT calculations as described in the text.}\label{dmj-tab-1}
\end{table}

\begin{table}
	\begin{tabular}{clllllll}
		Nucleus & $a_\text{iso}/$mT & $A_{xx}^\text{aniso}$/mT & $A_{yy}^\text{aniso}$/mT & $A_{zz}^\text{aniso}$/mT & $A_{xy}^\text{aniso}$/mT & $A_{xz}^\text{aniso}$/mT & $A_{yz}^\text{aniso}$/mT  \\
		\hline
		H1 & 2.308839 & 0.018394 & 0.005750 & -0.024144 & 0.119167 & -0.090257 & -0.105530 \\ 
		H2 & 0.903770 & -0.030255 & 0.134767 & -0.104512 & 0.111178 & 0.039520 & 0.065691 \\ 
		H3 & -0.034042 & 0.041327 & -0.039294 & -0.002033 & 0.017961 & 0.078922 & 0.025615 \\ 
		H4 & -0.077575 & 0.065617 & -0.016154 & -0.049462 & 0.036655 & 0.014217 & 0.004047 \\ 
		H5 & 1.071863 & 0.069089 & -0.054902 & -0.014187 & 0.013749 & -0.075976 & -0.006477 \\ 
		H6 & 0.258828 & 0.098308 & -0.041108 & -0.057200 & -0.024641 & 0.013959 & 0.002803 \\ 
		H7 & 2.308288 & 0.017844 & 0.006183 & -0.024028 & -0.119099 & -0.090068 & 0.105661 \\ 
		H8 & 0.902293 & -0.030775 & 0.135406 & -0.104631 & -0.110876 & 0.039322 & -0.065607 \\ 
		H9 & -0.034202 & 0.041235 & -0.039174 & -0.002061 & -0.018150 & 0.078901 & -0.025838 \\ 
		H10 & -0.077648 & 0.065415 & -0.015957 & -0.049458 & -0.036874 & 0.014222 & -0.004080 \\ 
		H11 & 1.073569 & 0.069102 & -0.054901 & -0.014201 & -0.014035 & -0.075981 & 0.006618 \\ 
		H12 & 0.259878 & 0.098464 & -0.041245 & -0.057219 & 0.024346 & 0.014054 & -0.002814 \\ 
		H13 & -0.166563 & 0.036159 & -0.000260 & -0.035899 & 0.038259 & -0.007026 & -0.004047 \\ 
		H14 & -0.166563 & 0.036159 & -0.000260 & -0.035899 & 0.038259 & -0.007026 & -0.004047 \\ 
		H15 & -0.166563 & 0.036159 & -0.000260 & -0.035899 & 0.038259 & -0.007026 & -0.004047 \\ 
		H16 & -0.166487 & 0.035983 & -0.000104 & -0.035879 & -0.038338 & -0.007021 & 0.004066 \\ 
		H17 & -0.166487 & 0.035983 & -0.000104 & -0.035879 & -0.038338 & -0.007021 & 0.004066 \\ 
		H18 & -0.166487 & 0.035983 & -0.000104 & -0.035879 & -0.038338 & -0.007021 & 0.004066 \\ 
		N1 & 0.831260 & -0.772676 & -0.781100 & 1.553776 & -0.000000 & -0.061480 & -0.000443
	\end{tabular}
	\caption{Hyperfine coupling parameters for the DMJ${}^{\bullet +}$ radical in the anti conformation, as labelled above. Isotropic components, $a_\text{iso} = (A_{xx}+A_{yy}+A_{zz})/3$, and anisotropic components, $A_{\alpha\beta}^\text{aniso} = A_{\alpha\beta}-\delta_{\alpha\beta}a_\text{iso}$, in the isolated radical principal axis frame are obtained from DFT calculations as described in the text.}\label{dmj-tab-2}
\end{table}

\subsection{Dipolar coupling and $g$ tensors}

The isotropic component of the $g$-tensor for the NDI${}^{\bullet -}$ radical is assumed to be the same as that for the \textit{N,N}-dipentyl NDI${}^{\bullet -}$ radical\cite{Andric2004}. The anisotropic components are obtained from DFT just as the anisotropic components of the hyperfine tensors are. The isotropic and anisotropic components of the $g$-tensors for DMJ${}^{\bullet +}$ are obtained from DFT in same way as the hyperfine coupling tensors. The same values are used for both anti and syn conformations, because the differences were found to be negligible. The off-diagonal elements for $g$ tensors in the principal axis frame of the isolated radicals are found to be small and for simplicity are set to zero. $g$ tensor parameters used are summarised in Table \ref{g-tab}. Overall it was found that the simulations are not sensitive to value of the $g$ tensors, due to the relatively low field strengths at which simulations are performed.

\begin{table}
	\begin{tabular}{cccccc}
		Radical & $g_\text{iso}$ & $g_{xx}-g_\text{iso}$ & $g_{yy}-g_\text{iso}$ & $g_{zz}-g_\text{iso}$   \\
		\hline
		NDI${}^{\bullet -}$ & 2.0040 & 0.0010 & 0.0007 & -0.0020 \\
		DMJ${}^{\bullet +}$ & 2.0031 & 0.0006 & 0.0001 & -0.0009
	\end{tabular}
	\caption{$g$ tensor components used in our simulations for the radicals, with $g_\text{iso}=(g_{xx}+g_{yy}+g_{zz})/3$.}\label{g-tab}
\end{table}

The dipolar coupling tensor was calculated based on the point dipole approximation, assuming the two radicals lie on the molecular $z$ axis exactly. The isotropic value of the electron g tensors were sued to calculate this coupling tensor. The radical separations used were $r = 16.5\ \text{\AA}$ for the n=1 molecule and $r = 20.9\ \text{\AA}$ for the n=2 molecule.\cite{Scott2009a}

\subsection{Rotational diffusion parameters}

\begin{figure}[h]
	\includegraphics[width=0.75\textwidth]{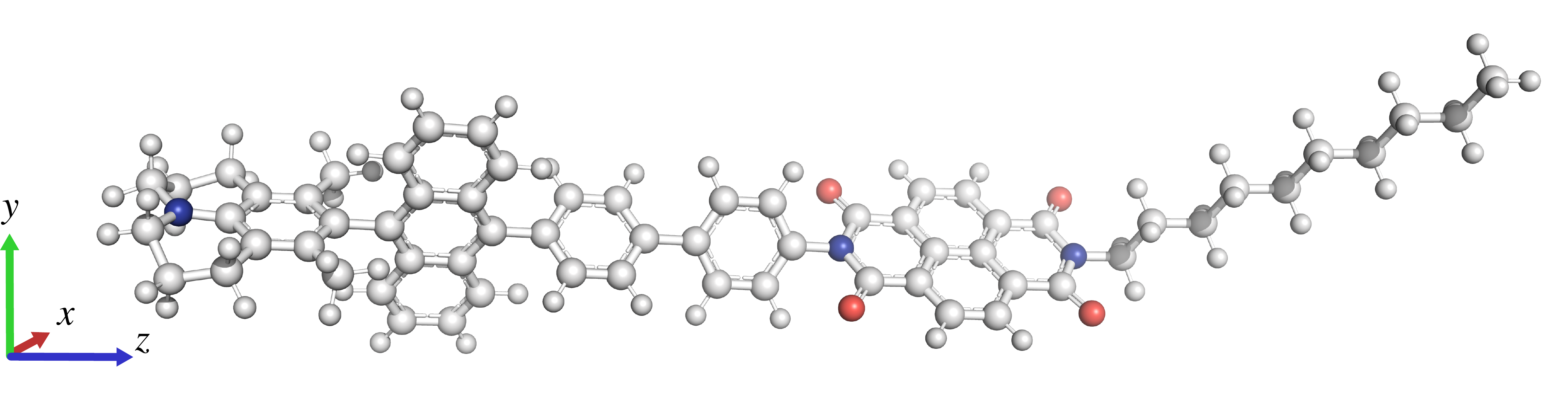}
	\caption{Example ground state geometry of the n=2 molecule with syn conformation of the DMJ${}^{\bullet+}$, calculated at the DFT B3LYP/6-31G(d,p) level of theory with D3 dispersion correction. Also shown is the approximate position of the molecular axes.}\label{mol-fig}
\end{figure}

The rotational motion of the molecule was simplified by treating it as a rigid prolate symmetric top. The isolated DMJ${}^{\bullet +}$ radical $y$ axis was taken to be parallel with the molecular $-z$ axis, and the molecular $x$ axis was taken lie on the radical $x$ axis (see Fig. \ref{mol-fig} for the n=2 case). The NDI${}^{\bullet -}$ $x$ axis was taken to lie on the molecular $z$ axis, and the isolated radical $y$ axis was taken to be at an angle $\theta$ from the molecular $x$ axis. This angle for the solution phase radical ion pair state was estimated from the gas phase groundstate geometry, optimized at the DFT B3LYP/6-31G(d,p) level of theory with D3 dispersion correction. For n=1 the angle was taken to be $-82^\circ$ and for n=2 the angle was taken to be $-18^\circ$. This angle is independent of whether the DMJ${}^{\bullet+}$ is in the syn or anti conformation.

The molecule was assumed to behave as a rigid rotor, with rotational diffusion constants given by the Stokes-Einstein equation,
\begin{align}
D_\alpha = \frac{k_\mathrm{B}T}{8\pi \eta r_\alpha^3}
\end{align}
where $T$ is the temperature, 295K,\cite{Scott2009a}, $\eta$ is the bulk dynamic viscosity, taken to be 0.5812 mPa s, and $r_\alpha$ is the hydrodynamic radius of the molecule in the $\alpha=\parallel\text{ or }\perp$ direction. We estimated $r_\parallel$ as 8.50 \AA, and $ r_\perp$  as 15.45 \AA\ and 17.50 \AA\ for n=1 and n=2 molecules respectively. These estimates are based on atom-to-atom distances in the ground state gas phase geometry, adding the van der Waals radii of hydrogen of 1.2 \AA\ and 2.8 \AA\ for the solvent toluene,\cite{Schulze2014} ignoring the n-octyl group on the NDI, and as such these are likely to be underestimates of the true hydrodynamic radii. 

However, the Stokes-Einstein equation is also only approximate in this case given that the molecular volume of toluene is not significantly smaller than that of the molecule, and as a result the viscosity experienced by the molecule may be different to that of the bulk. Furthermore the two radicals will be able to rotate relative to each other, i.e. the moelcule is not a true rigid body. Given these approximations, we have tested the sensitivity of the fits and fitted parameters of our models, increasing and decreasing the diffusion constants by $50\%$. The fits and fitted parameters were found to not change significantly on altering the diffusion constants in this way.

%
%

\bibliography{si.bib}